\documentclass[fleqn,10pt,a4paper]{article}
\usepackage{amsmath,amssymb,authblk,bm}
\usepackage{cite,epsfig,color,graphicx,subfigure}
\title{Quantum backreaction in laser-driven plasma}

\author[3, 1]{A. Conroy}
\author[1]{C. Fiedler}
\author[2]{A. Noble}
\author[1]{D. A. Burton}
\affil[1]{Department of Physics\\ Lancaster University\\ Lancaster\\ LA1 4YB\\ UK}
\affil[2]{Department of Physics\\ SUPA and University of Strathclyde\\ Glasgow\\ G4 0NG\\ UK}
\affil[3]{Centre for Astrophysics and Relativity\\ School of Mathematical Sciences\\\newline Dublin City University\\ Glasnevin\\ Dublin 9\\ Rep. of Ireland} 
\begin{document}
\maketitle
\begin{abstract}
We present a new approach for investigating quantum effects in laser-driven plasma. Unlike the modelling strategies underpinning particle-in-cell codes that include the effects of quantum electrodynamics, our new field theory incorporates multi-particle effects from the outset. Our approach is based on the path-integral quantisation of a classical bi-scalar field theory describing the behaviour of a laser pulse propagating through an underdense plasma. Results established in the context of quantum field theory on curved spacetime are used to derive a non-linear, non-local, effective field theory that describes the evolution of the laser-driven plasma due to quantum fluctuations. As the first application of our new theory, we explore the behaviour of perturbations to fields describing a uniform, monochromatic, laser beam propagating through a uniform plasma.
Our results suggest that quantum fluctuations could play a significant role in the evolution of an underdense plasma driven by an x-ray laser pulse.
\end{abstract}

\section{Introduction}
\label{sec:introduction}
The new generation of high-power laser systems will drive the experimental study of high-intensity laser-matter interactions into novel territory. Forthcoming facilities~\cite{eli} are expected to allow experimental investigations of uncharted parameter regimes using laser pulses with unprecedented peak intensities (greater than $10^{22}\,\rm{W}\rm{cm}^{-2}$). In such regimes, the relativistic quantum-mechanical aspects of laser-matter interactions must be included~\cite{dipiazza:2012, burton:2014}, and this requirement has driven the development of particle-in-cell codes, such as EPOCH~\cite{arber:2015}, that incorporate the effects of quantum electrodynamics. In such codes, the matter is represented by a large number of classical macro-particles whose electromagnetic fields, and the laser field, serve as a background in the perturbative calculation of single-particle matrix elements associated with each macro-particle. Each quantum process is assumed to be active only within a spacetime region whose size is negligible in comparison to the classical length and time scales of the laser and matter variables. Interactions between macro-particles in this model are implemented through their contributions to Maxwell equations as classical sources; as a consequence, the multi-particle effects calculated using this approach are classical, rather than quantum, in origin. However, multi-particle, collective, effects typically dominate the behaviour of classical laser-driven plasma, and strategies for modelling quantum effects based on single-particle considerations may not be adequate in this context. Unfortunately, an ab-initio multi-particle quantum treatment of the detailed behaviour of a laser-driven plasma is a formidable task. Even ab-initio simulation tools based solely on classical laser-plasma physics require high-performance computing facilities to be of greatest utility, and the computational demands of a fully quantum calculation are substantially heavier. We are not aware of any computer hardware or software suitable for undertaking an ab-initio numerical analysis of the multi-particle quantum behaviour of a laser-driven plasma, and so it is vital to undertake a thorough investigation of all suitable alternatives.  

Even without quantum effects, the 3-dimensional calculation of the propagation of micrometre-length laser pulses through many centimetres of underdense plasma is a considerable computational challenge. Such considerations have motivated the development of models of classical laser-driven plasma with reduced degrees of freedom, such as those underpinning INF\&RNO~\cite{benedetti:2018}, leading to a reduction of the computational burden by several orders of magnitude. The vigorous effort devoted to the development of such computational tools shows no sign of abating; however, the role of quantum theory in the context of reduced models is yet to be thoroughly investigated. Moreover, a quantum theory based on a reduced model may capture important physics inaccessible to classical particle-in-cell codes augmented by single-particle quantum processes only.

This article offers a new theory for investigating the quantum effects exhibited by a laser pulse propagating through an underdense plasma. Rather than modelling the microscopic degrees of freedom by appealing to quantum electrodynamics, our approach is based on the quantisation of a fluid description of the laser-driven plasma. Hence, multi-particle considerations are included from the outset. The underpinning ingredients rely on standard approximations in classical laser-plasma theory; the slowly-varying envelope approximation for the laser pulse, and the ponderomotive approximation for the force exerted by the laser on the plasma electrons~\cite{esarey:2009}. We exploit results established in the context of quantum field theory on curved spacetime to obtain the quantum corrections to the classical field equations. However, it is worth noting that one can also motivate our approach from considerations based on scalar QED, as shown in Appendix~\ref{appendix:QED}. It is clear that one cannot dispense with either the laser or matter content without ruining the connection between our theory and scalar QED; thus, our theory is not applicable in the vacuum limit.

As the first application of our new theory, we show that the quantum fluctuations lead to the relationship
\begin{equation}
\label{ultrarel_disp_final_intro}
\omega = v\kappa + \sqrt{\frac{3\alpha}{128\,\pi^2}}\frac{\lambda_e}{w_0}\bigg[\frac{a_0^2}{(a_0^2+1)^3}\bigg]^{\frac{1}{4}}\frac{c^2 \kappa^2}{\sqrt{\omega_0 \omega_p}}
\end{equation}
between the angular frequency $\omega$ and wavenumber $\kappa$ of a longitudinal perturbation to the laser-plasma variables in the rest frame of the plasma ions. The quantity $a_0$ is the dimensionless amplitude of the laser pulse, $\omega_0$ is the frequency of the laser pulse, $\omega_p$ is the plasma frequency in the unperturbed state, $\lambda_e$ is the Compton wavelength of the electron, $\alpha$ is the fine-structure constant, $w_0$ is the width of the laser pulse, and the speed $v$ satisfies
\begin{equation}
\label{ultrarel_phase_speed_intro}
v = \frac{\omega_p}{\omega_0}\frac{a_0 c}{2\sqrt{a_0^2+1}} + {\cal O}(\omega_0^{-2}).
\end{equation}
Equations (\ref{ultrarel_disp_final_intro}), (\ref{ultrarel_phase_speed_intro}) are valid for perturbations along the axis of the pulse whose wavelength is shorter than the length of the pulse.

Using (\ref{ultrarel_disp_final_intro}), the characteristic time scale $\tau$ over which the length of a Gaussian wave packet increases by the fractional amount $\delta$ due to quantum fluctations, in the rest frame of the plasma ions, satisfies
\begin{equation}
\label{ultrarel_time_final_intro}
\tau \sim \sqrt{\frac{128\,\pi^2}{3\alpha}}\frac{w_0}{\lambda_e}\bigg[\frac{(a_0^2+1)^3}{a_0^2}\bigg]^{\frac{1}{4}}\frac{\sqrt{\omega_0 \omega_p}}{c^2}\,\sigma^2\frac{\sqrt{(1+\delta)^2-1}}{2},
\end{equation}
where $\sigma$ is the initial length of the wave packet. The implications of (\ref{ultrarel_time_final_intro}) are readily appreciated by expressing it in terms of the quantities $\check\tau$, $\check\sigma$ normalised with respect to the laser period $2\pi/\omega_0$ and laser wavelength $\lambda_0 = 2\pi c/\omega_0$, respectively. Hence
\begin{equation}
\label{ultrarel_time_normalised_final_intro_1}
\check\tau \sim 1509\bigg[\frac{(a_0^2+1)^3}{a_0^2}\bigg]^{\frac{1}{4}}\sqrt{\frac{w_0^2\lambda_0}{\lambda_e^2\lambda_p}}\,\check\sigma^2\frac{\sqrt{(1+\delta)^2-1}}{2} 
\end{equation}
where the length $\lambda_p = 2\pi c/\omega_p$ has been introduced. Moreover, the theory underpinning (\ref{ultrarel_disp_final_intro}) requires the laser wavelength $\lambda_0$ to be the shortest classical length scale in the analysis; thus, $\check\sigma > 1$ and
 \begin{equation}
\label{ultrarel_time_normalised_final_intro_2}
\check\tau \gtrsim 1509\bigg[\frac{(a_0^2+1)^3}{a_0^2}\bigg]^{\frac{1}{4}}\sqrt{\frac{w_0^2\lambda_0\delta}{2\lambda_e^2\lambda_p}} 
\end{equation}
follows immediately when $\delta\ll 1$.

Experimental facilities are available in which intense laser pulses propagate through an underdense plasma over distances that are many multiples of the classical Rayleigh length. A comparison of (\ref{ultrarel_time_normalised_final_intro_2}) with the number of oscillations $N\pi w_0^2/\lambda_0^2$ corresponding to $N$ multiples of the Rayleigh length of a laser beam with waist $w_0$ yields the upper bound
\begin{equation}
\label{upper_bound_delta}
\delta \lesssim 8.7\times 10^{-6}\frac{a_0}{(a_0^2+1)^{3/2}}\frac{N^2 w_0^2 \lambda_e^2 \lambda_p}{\lambda_0^5}
\end{equation}
on the corresponding fractional change $\delta$ in the length of the Gaussian wave packet.

Inspection of (\ref{upper_bound_delta}) suggests that dispersive effects due to quantum fluctuations will not be detectable in any contemporary experiment based on an underdense plasma driven by an optical laser. For example, the parameter choice $w_0 = \lambda_p = 30\,{\rm \mu m}$ is suitable for maintaining an intense optical laser pulse with $\lambda_0=800\,{\rm nm}$ over tens of Rayleigh lengths (for general considerations, see Ref.~\cite{esarey:2009}). Even though the maximum value of (\ref{upper_bound_delta}) in $a_0$ ($a_0 \approx 0.7$) is commonly achieved in experiments using high-power optical lasers, such parameters yield estimates for $\delta$ (e.g. $\delta \lesssim 2.6\times 10^{-9}$ when $N=40$) that are unresolvable in any realistic experiment. However, the strong dependence of (\ref{upper_bound_delta}) on $\lambda_0$ suggests that x-ray lasers may be a better prospect, even though their intensity is considerably lower than that achieved by their optical brethren. For example, an x-ray pulse with $\lambda_0=10\,{\rm nm}$ and $w_0 = 100\,{\rm \mu m}$, propagating through a plasma with the matched length $\lambda_p = 100\,{\rm \mu m}$, gives an upper bound on $\delta$ of approximately $5\%$ when $a_0 = 6 \times 10^{-5}$ ($\sim 10^{14}\,{\rm W}\,{\rm cm}^{-2}$) and $N=40$. Thus, a comprehensive experimental investigation of our new results may be possible using an x-ray laser (e.g. the European XFEL~\cite{xfel}) with pulses of duration greater than approximately $350\,{\rm fs}$.

Section~\ref{sec:classical_theory} introduces the classical theory underpinning our approach. Section~\ref{sec:quantum_theory} details the 1-loop effective action that arises from a path-integral quantisation of the underlying classical theory, and Section~\ref{sec:field_equations} summarises the non-linear field equations that emerge. By construction, the field equations include the quantum backreaction of the laser-plasma system. Section~\ref{sec:linearised_field_equations} is a perturbative analysis of the field equations, applicable when the wavelength of the perturbation to the laser-plasma variables is much shorter than the length of the laser pulse. Equation (\ref{ultrarel_disp_final_intro}) emerges as a result.
\section{Classical theory of laser-driven plasma}
\label{sec:classical_theory}
Our particular interest here is in the interaction of electrons with an intense laser pulse propagating through an underdense plasma, where the internal oscillations of the pulse determine the shortest significant classical length and time scales of the system. The large difference between the scales associated with the internal oscillations of the laser pulse and the behaviour of the wake behind the front of the pulse permits approximations to be introduced that greatly simplify the analysis. The precise details of the plasma electron motion due to the fast oscillations of the fields within the laser pulse are sacrificed to obtain an efficient model of the electron dynamics over distances that are much greater than the wavelength of the laser. Computationally efficient models in this context typically exploit the ponderomotive approximation for calculating the effect of the laser pulse on the plasma electrons in tandem with a slowly-varying envelope approximation for determining the influence of the matter on the laser pulse~\cite{benedetti:2018}.

The total electric field and total magnetic field are each expressed a sum of two terms. The first term can be understood as the local average of the respective total field over the fast oscillations of the laser pulse, and the second term is the field of the laser itself. In particular, the total electric field ${\bf E}_{\rm tot}$ satisfies ${\bf E}_{\rm tot} = {\bf E} - \partial_t {\bf A}_0$ where, for notational convenience, the local average of ${\bf E}_{\rm tot}$ is denoted ${\bf E}$ and the laser pulse is encoded by the vector potential ${\bf A}_0$. Likewise, the total magnetic field ${\bf B}_{\rm tot}$ is ${\bf B}_{\rm tot} = {\bf B} + \bm{\nabla} \times {\bf A}_0$.

The electrons in a cold fluid model of a laser-plasma satisfy
\begin{align}
\label{momentum_balance}
\partial_t{\bf p} + ({\bf v}\cdot\bm{\nabla}){\bf p} = -\frac{e^2}{2 m_e \gamma}\bm{\nabla}\langle{\bf A}_0^2\rangle - e({\bf E} + {\bf v}\times {\bf B}),\quad\gamma=\sqrt{1+\frac{{\bf p}^2}{m_e^2 c^2} + \frac{e^2\langle{\bf A}_0^2\rangle}{m_e^2 c^2}},\quad {\bf p} = m_e \gamma\,{\bf v}
\end{align}
in the relativistic ponderomotive approximation, where $m_e$ is the rest mass of the electron, $e$ is the elementary charge and $c$ is the speed of light in vacuo. The effect of the laser pulse on the electrons is determined by $\langle {\bf A}_0^2\rangle$, the square of the magnitude of the vector potential ${\bf A}_0$ of the pulse averaged over its fast internal oscillations. The vector fields ${\bf v}$, ${\bf p}$ are understood as the averaged velocity and averaged momentum, respectively, of the plasma electrons. The contribution to $\gamma^2$ (the square of the Lorentz factor $\gamma$) proportional to $\langle {\bf A}_0^2\rangle$ is due to the fast oscillatory motion of the electrons induced by the laser pulse.

The fields ${\bf E}$, ${\bf B}$ are produced from the averaged properties of the electrons; in particular,
\begin{align}
\label{maxwell}
&\bm{\nabla}\cdot{\bf D} = - en + \rho_i,\quad\bm{\nabla}\times{\bf H} = \partial_t{\bf D} - en{\bf v} + {\bf j}_i,\quad  \bm{\nabla}\times{\bf E} = - \partial_t {\bf B},\quad \bm{\nabla}\cdot{\bf B} = 0,
\end{align}
with ${\bf D} = \varepsilon_0 {\bf E}$, ${\bf H} = {\bf B}/\mu_0$, where $\varepsilon_0$, $\mu_0$ are the permittivity and permeability of the vacuum, respectively, and $n$ is the averaged electron number density. The fields $\rho_i$, ${\bf j}_i$ are the ion charge density and ion current density, respectively, and are specified as data. 

To proceed further, the properties of the laser potential ${\bf A}_0$ must be specified. A popular strategy used in many studies of laser-plasma accelerators~\cite{esarey:2009} is to solve
\begin{equation} 
\label{laser_PDE}
\partial_t^2 {\bf A}_0 - c^2 \bm{\nabla}^2{\bf A}_0 = -\omega_p^2{\bf A}_0,\qquad \omega_p=\sqrt{\frac{e^2\,n}{\varepsilon_0 m_e \gamma}}
\end{equation}
for the vector potential ${\bf A}_0$, where $\omega_p$ is the plasma frequency given by the local electron number density $n$. For practical purposes, slowly-varying envelope, or eikonel, approximations are commonly used to remove the fast oscillations from (\ref{laser_PDE}) before further analysis. In addition to the separation of scales, a complete justification of the model (\ref{momentum_balance}), (\ref{maxwell}), (\ref{laser_PDE}) requires the dominant component of the electron momentum to be parallel to the direction of propagation of the laser pulse. We will exploit this facet of the classical model when developing the quantum theory in Section~\ref{sec:quantum_theory}.
 
It can be shown~\cite{cairns:2004} that applying the eikonel approximation to (\ref{laser_PDE}) yields the conservation of wave action 
\begin{equation}
\label{wave_conservation}
\partial_t (\langle {\bf A}_0^2\rangle\,\omega_0) + c^2\,\bm{\nabla}\cdot(\langle {\bf A}_0^2\rangle{\bf k}_0)=0,
\end{equation}
where the local frequency $\omega_0$ and local wave vector ${\bf k}_0$ of the laser pulse satisfy $\partial_t {\bf k}_0 = -\bm{\nabla}\omega_0$ and the local dispersion relation
\begin{equation}
\label{dispersion}
\omega_0^2 - c^2{\bf k}_0^2 = \omega^2_p.
\end{equation}
Equations (\ref{momentum_balance}), (\ref{maxwell}), (\ref{wave_conservation}), (\ref{dispersion}) constitute a closed system of classical field equations for a laser-driven plasma.
\subsection{Reduction of the classical theory}
\label{sec:reduction_classical_theory}
There are numerous ways of developing effective quantum theories from the above system of classical field equations. A simple strategy is to focus on a regime in which the effects due to the interaction between the laser field and matter dominate over those directly connected to the averaged electromagnetic fields; thus, ${\bf E}$, ${\bf B}$ are treated as negligible. To lowest order, the field equation for ${\bf p}$ is (\ref{momentum_balance}) with ${\bf E}$, ${\bf B}$ set to zero:
\begin{equation}
\label{momentum_balance_no_fields}
\partial_t{\bf p} + ({\bf v}\cdot\bm{\nabla}){\bf p} = -\frac{e^2}{2 m_e \gamma}\bm{\nabla}\langle{\bf A}_0^2\rangle.
\end{equation}
Further simplification is achieved by focussing on potential flow. Equation (\ref{momentum_balance_no_fields}) can be expressed as
\begin{equation}
\partial_t{\bf p} - {\bf v}\times(\bm{\nabla}\times{\bf p}) = -m_e c^2\bm{\nabla}\gamma,
\end{equation}
which is solved by ${\bf p}=\bm{\nabla}\widetilde\Psi$ with the momentum potential $\widetilde\Psi$ satisfying $\partial_t \widetilde\Psi = - m_e c^2 \gamma$. The latter can be rearranged to give
\begin{equation}
\label{A_l_psi}
e^2 c^2 \langle{\bf A}_0^2\rangle = (\partial_t \widetilde\Psi)^2 - c^2(\bm{\nabla}\widetilde\Psi)^2 - m_e ^2 c^4.
\end{equation}

Although ${\bf E}$, ${\bf B}$ feature in the lowest order Maxwell equations (\ref{maxwell}), their precise forms are not needed; the only necessary consequence of (\ref{maxwell}) is charge conservation. The electron number density $n$ is required to obtain the plasma frequency $\omega_p$, and charge conservation provides a suitable field equation for $n$. Since the ion charge density is locally conserved, the electron number density must satisfy
\begin{equation}
\label{charge_conservation}
\partial_t n + \bm{\nabla}\cdot(n{\bf v}) = 0
\end{equation}
which, using the expression for $\omega_p$ given in (\ref{laser_PDE}), yields
\begin{equation}
\label{psi_PDE}
\partial_t(\omega_p^2\,\partial_t\widetilde\Psi) - c^2 \bm{\nabla}\cdot(\omega_p^2\,\bm{\nabla}\widetilde\Psi) = 0.
\end{equation}
where $m_e {\bf v} = \bm{\nabla}\widetilde\Psi/\gamma$, $\partial_t \widetilde\Psi = - m_e c^2 \gamma$ have been employed.

The quantity $\varepsilon_0 \omega_p^2 \langle{\bf A}_0^2\rangle/2$ has the physical dimensions of an energy density, and it can serve as the Lagrangian density in an action principle for the field equations (\ref{wave_conservation}), (\ref{psi_PDE}). Expressing $\varepsilon_0 \omega_p^2 \langle{\bf A}_0^2\rangle/2$ in terms of $\widetilde\Psi$ and the phase $\widetilde\Phi$, where $\omega_0=-\partial_t \widetilde\Phi$, ${\bf k}_0 = \bm{\nabla}\widetilde\Phi$, suggests the action
\begin{equation}
\label{3+1_action}
{\cal S}[\widetilde\Phi,\widetilde\Psi] = \frac{\varepsilon_0\sigma_*}{e^2 c^2}\int dt d^3\bm{x}\,\frac{1}{2}\big\{(\partial_t \widetilde\Phi)^2 - c^2(\bm{\nabla}\widetilde\Phi)^2\big\}\big\{(\partial_t \widetilde\Psi)^2 - c^2(\bm{\nabla}\widetilde\Psi)^2 - m_e^2 c^4\big\}
\end{equation}
where (\ref{dispersion}), (\ref{A_l_psi}) have been used to substitute $\omega_p$, $\langle{\bf A}_0^2\rangle$, respectively. Stationary variations of (\ref{3+1_action}) with respect to $\widetilde\Phi$, $\widetilde\Psi$ yield (\ref{wave_conservation}), (\ref{psi_PDE}), respectively. The dimensionless constant $\sigma_*$ has been introduced because (\ref{3+1_action}) has been obtained using dimensional reasoning and, although $\sigma_*$ is inert in the classical theory, it will scale the quantum corrections to the classical field equations.

Although substantial simplifications have been made to obtain (\ref{3+1_action}), it is highly beneficial, from the perspective of quantum theory, to replace (\ref{3+1_action}) with its counterpart theory in one spatial dimension. This strategy is physically justified by noting that the laser-plasma variables $\widetilde\Phi$, $\widetilde\Psi$ change over a much shorter distance parallel to the direction of propagation of the laser pulse than transverse to it. This property of the laser-plasma system is closely connected to the justification behind the introduction of (\ref{momentum_balance}), (\ref{laser_PDE}). Furthermore, within this approximation, it is reasonable to choose $\omega_p^2 \langle{\bf A}_0^2\rangle$ to be expressible as a product of a function of $(t,z)$ and a function of $(x,y)$. Thus, the quantity $L_*$ given by
\begin{equation}
L_* = \sqrt{\frac{\sigma_*\int dx dy\,\omega_p^2 \langle{\bf A}_0^2\rangle}{\omega_p^2 \langle{\bf A}_0^2\rangle|_{x=y=0}}}
\end{equation}
is constant, and we arrive at the estimate
\begin{equation}
\label{3+1_action_simplified}
{\cal S}[\widetilde\Phi,\widetilde\Psi] \approx \frac{\varepsilon_0 L_*^2}{e^2 c^2}\int dt dz\,\frac{1}{2}\big\{(\partial_t \widetilde\Phi)^2 - c^2(\partial_z\widetilde\Phi)^2\big\}\big\{(\partial_t \widetilde\Psi)^2 - c^2(\partial_z\widetilde\Psi)^2 - m_e^2 c^4\big\}|_{x=y=0}
\end{equation}
for the action (\ref{3+1_action}). The line $x=y=0$ has been chosen to lie along the centre of the laser pulse. 

The above considerations suggest a relativistic bi-scalar field theory on $2$-dimensional spacetime given by the action
\begin{equation}
\label{1+1_action}
S[\Phi,\Psi] = \hbar \int d^2 x \sqrt{-\eta}\,\frac{1}{2}\eta^{\mu\nu}\partial_\mu \Phi\partial_\nu \Phi \,(\eta^{\sigma\tau}\partial_\sigma \Psi\partial_\tau\Psi + 1)
\end{equation}
where $\eta_{\mu\nu}$ is the Minkowski metric with signature $(-,+)$, and $\mu, \nu = 0,1$. The fields $\Phi$, $\Psi$, $\eta_{\mu\nu}$, and coordinates $x^0$, $x^1$ are dimensionless. The action (\ref{1+1_action}) can be obtained from (\ref{3+1_action_simplified}) using the substitutions
\begin{equation}
\label{dimensionless_variables}
x^0 = \frac{c t}{l_*},\qquad x^1 = \frac{z}{l_*},\qquad \widetilde\Phi|_{x=y=0} = \sqrt{\frac{\hbar e^2}{\varepsilon_0 m_e^2 c^3 L_*^2}}\,\Phi,\qquad \widetilde\Psi|_{x=y=0} = m_e c l_* \Psi.
\end{equation}
Note that the length scales $l_*$, $L_*$ are unrelated; the former is inert (it has no direct physical meaning) and is introduced solely for mathematical elegance, whereas the latter is proportional to the transverse size (width) of the laser pulse. It is also possible to motivate (\ref{1+1_action}) using considerations based on scalar QED; see Appendix~\ref{appendix:QED}.

Unless otherwise indicated, for convenience we will henceforth adopt units in which the reduced Planck constant $\hbar$ is unity.
\section{Quantum considerations}
\label{sec:quantum_theory}
A perturbative exploration of some of the quantum implications of
\begin{equation}
\label{1+1_action_simp}
S[\Phi,\Psi] = \int d^2 x \sqrt{-\eta}\,\frac{1}{2}\eta^{\mu\nu}\partial_\mu \Phi\partial_\nu \Phi \,(\eta^{\sigma\tau}\partial_\sigma \Psi \partial_\tau\Psi + 1)
\end{equation}
can be undertaken using the $1$-loop effective action $\Gamma$ given by
\begin{equation}
\label{1-loop_action}
\Gamma[\vec\Phi] = S[\vec\Phi] - i\ln\bigg\{\int{\cal D}\vec f\,\exp(i\Lambda[\vec f])\bigg\}
\end{equation}
where
\begin{equation}
\label{lambda}
\Lambda[\vec f] =\frac{1}{2}\int d^2 x\int d^2 x^\prime\frac{\delta^2 S}{\delta \Phi_A(x)\delta\Phi_B(x^\prime)}f_A(x)f_B(x^\prime)
\end{equation} 
with the indices $A,B$ ranging over $1,2$ and $\Phi_1 = \Phi$, $\Phi_2 = \Psi$. For convenience,  the notation $\vec\Phi = (\Phi_1\; \Phi_2)^{\rm T}$, $\vec f = (f_1\; f_2)^{\rm T}$ has been introduced, with ${\rm T}$ denoting matrix transposition. The functional $\Lambda$ can be expressed as
\begin{align}
\notag
\Lambda[\vec f] =& \int d^2 x \sqrt{-\eta}\,\frac{1}{2}\big\{\big(\eta^{\mu\nu}\partial_\mu \Psi \partial_\nu \Psi + 1\big)\,\eta^{\sigma\tau}\partial_\sigma f_1 \partial_\tau f_1 + \eta^{\mu\nu}\partial_\mu \Phi\partial_\nu \Phi \,\eta^{\sigma\tau}\partial_\sigma f_2\partial_\tau f_2\\
\label{quadratic_action_two_comp}
&\qquad\qquad\qquad + 4\eta^{\mu\nu}\partial_\mu \Phi\partial_\nu f_1\,\eta^{\sigma\tau}\partial_\sigma \Psi\partial_\tau f_2\big\}
\end{align}
or, equivalently,
\begin{align}
\Lambda[\vec f] = \int d^2 x \sqrt{-\eta}\frac{1}{2}\vec{f}^{\,\dag} {\cal O} \vec{f}
\end{align}
using integration by parts, where $\vec{f}^{\,\dag}$ is the Hermitian conjugate of $\vec{f}$. In the above, and throughout the following, we adopt the minimal approach in which $\vec f$ is regarded as a map on an arbitrarily large torus; hence, boundary terms do not arise when integration by parts is used. The operator ${\cal O}$ is given by
\begin{align}
\label{O_def}
{\cal O}\vec f = - \nabla^{(\eta)}_\mu
\bigg\{
\begin{pmatrix}
(\eta^{\sigma\tau}\partial_\sigma \Psi \partial_\tau \Psi + 1)\eta^{\mu\nu} & 2\eta^{\mu\sigma}\partial_\sigma \Phi\,\eta^{\nu\tau}\partial_\tau \Psi \\
2\eta^{\mu\sigma}\partial_\sigma \Psi\,\eta^{\nu\tau}\partial_\tau \Phi & \eta^{\sigma\tau}\partial_\sigma \Phi \partial_\tau \Phi\,\eta^{\mu\nu}\\
\end{pmatrix}
\begin{pmatrix}
\partial_\nu f_1 \\
\partial_\nu f_2
\end{pmatrix}
\bigg\}
\end{align}
with $\nabla^{(\eta)}_\mu$ the Levi-Civita covariant derivative induced from $\eta_{\mu\nu}$. Since ${\cal O}$ is Hermitian with respect to the inner product $(\vec{a}, \vec{b}) = \int d^2 x \sqrt{-\eta}\, \vec{a}^{\,\dag} \vec{b}$, ${\cal O}$ has real eigenvalues. By definition, $\int{\cal D}\vec f\,\exp(i\Lambda[\vec f]) = 1/\sqrt{{\rm det}(-i{\cal O})}$ where the functional determinant ${\rm det}(-i{\cal O})$ is formally equal to the product $\Pi_n (-i\lambda_n)$ of the non-zero eigenvalues $\{-i\lambda_n\}$ of $-i{\cal O}$.

In general, it is not straightforward to analytically compute ${\rm det}(-i{\cal O})$. However, the calculation is trivial when $\Phi$, $\Psi$ are linear functions of Minkowski coordinates adapted to $\eta_{\mu\nu}$ because, in that case, the matrix within the curly brackets in (\ref{O_def}) is constant when expressed in those coordinates. Thus, the eigenfunctions of ${\cal O}$ have the form $(a\; b)^{\rm T}\exp(i l_\mu x^\mu)$, where the components of the wave $2$-vector $l_\mu$ and the coefficients $a$, $b$ are constant, and the pair of eigenvalues $\lambda^+_{\vec l}$, $\lambda^-_{\vec l}$ corresponding to each $l_\mu$ satisfies
\begin{align}
\label{lambda_prod}
\lambda^+_{\vec l} \lambda^-_{\vec l} = (\eta^{\sigma\tau}\partial_\sigma \Psi \partial_\tau \Psi + 1)\eta^{\gamma\delta}\partial_\gamma \Phi \partial_\delta \Phi\,(\eta^{\mu\nu} l_\mu l_\nu)^2 - 4 (\eta^{\mu\sigma}\partial_\sigma \Phi\,\eta^{\nu\tau}\partial_\tau \Psi\, l_\mu l_\nu)^2.
\end{align}
For subsequent analysis, the fact that the right-hand side of (\ref{lambda_prod}) can be readily factorised is key. It follows that
\begin{equation}
\label{lambda_prod_A+A-}
\lambda^+_{\vec l} \lambda^-_{\vec l} = ({\cal A}^{\mu\nu}_+ l_\mu l_\nu)({\cal A}^{\sigma\tau}_- l_\sigma l_\tau)
\end{equation}
where the symmetric tensors ${\cal A}^{\mu\nu}_+$, ${\cal A}^{\mu\nu}_-$ are
\begin{align}
\label{calA_+_def}
{\cal A}^{\mu\nu}_+ &= s\bigg\{\sqrt{\big(\eta^{\sigma\tau}\partial_\sigma \Psi \partial_\tau \Psi + 1\big)\eta^{\gamma\delta}\partial_\gamma \Phi \partial_\delta \Phi}\,\eta^{\mu\nu} + \eta^{\mu\sigma}\eta^{\nu\tau}\partial_\sigma \Phi \partial_\tau \Psi + \eta^{\mu\sigma}\eta^{\nu\tau}\partial_\sigma \Psi \partial_\tau \Phi\bigg\},\\
\label{calA_-_def}
{\cal A}^{\mu\nu}_- &= s\bigg\{\sqrt{\big(\eta^{\sigma\tau}\partial_\sigma \Psi \partial_\tau \Psi + 1\big)\eta^{\gamma\delta}\partial_\gamma \Phi \partial_\delta \Phi}\,\eta^{\mu\nu} - \eta^{\mu\sigma}\eta^{\nu\tau}\partial_\sigma \Phi \partial_\tau \Psi - \eta^{\mu\sigma}\eta^{\nu\tau}\partial_\sigma \Psi \partial_\tau \Phi\bigg\}
\end{align}
and the constant $s$ satisfies $s^2=1$. The presence of $s$ reflects some of the freedom in the solution to (\ref{lambda_prod_A+A-}), although we will find that $s=-1$ emerges as a consequence of the analysis. Although there is freedom to scale ${\cal A}^{\mu\nu}_+$, ${\cal A}^{\mu\nu}_-$ whilst leaving the product ${\cal A}^{\mu\nu}_+{\cal A}^{\sigma\tau}_-$ invariant, we will not explore that possibility here. Equations (\ref{calA_+_def}), (\ref{calA_-_def}) are the natural choice in the present context. 

It then follows ${\rm det}(-i{\cal O}) =  \Pi_{\vec l} (i\lambda^+_{\vec l} i\lambda^-_{\vec l}) = \Pi_{\vec l} (-i\lambda^+_{\vec l})\,\Pi_{\vec q}(-i\lambda^-_{\vec q}) = {\rm det}(-i{\cal O}_+)\,{\rm det}(-i{\cal O}_-)$ where the second-order differential operators ${\cal O}_+$, ${\cal O}_-$ are
\begin{equation}
\label{O_pm}
{\cal O_+} f = -  \nabla^{(\eta)}_\mu ({\cal A}^{\mu\nu}_+ \partial_\nu f),\qquad{\cal O_-} f = -  \nabla^{(\eta)}_\mu ({\cal A}^{\mu\nu}_- \partial_\nu f).
\end{equation}
Hence, one can factorise $\int {\cal D}{\vec f}\exp(i\Lambda[\vec f])$ as
\begin{equation}
\label{Lambda_factor}
\int {\cal D}{\vec f}\exp(i\Lambda[\vec f]) = \bigg\{\int {\cal D}f\exp(i\Lambda_+[f])\bigg\}\bigg\{\int {\cal D}f\exp(i\Lambda_-[f])\bigg\}
\end{equation}
where
\begin{equation}
\label{Lambda_pm}
\Lambda_+[f] = \int d^2 x \sqrt{-\eta} \frac{1}{2}{\cal A}_+^{\mu\nu}\partial_\mu f \partial_\nu f,\qquad \Lambda_-[f] = \int d^2 x \sqrt{-\eta} \frac{1}{2}{\cal A}_-^{\mu\nu}\partial_\mu f \partial_\nu f.
\end{equation}

The above considerations are strictly only applicable to the cases where $\Phi$, $\Psi$ are linear functions of Minkowski coordinates adapted to $\eta_{\mu\nu}$. Fields $\Phi$, $\Psi$ of this type describe a non-evolving monochromatic laser beam propagating through a uniform plasma. However, it is plausible that (\ref{calA_+_def}), (\ref{calA_-_def}), (\ref{Lambda_factor}), (\ref{Lambda_pm}) hold to a reasonable approximation when $\Phi$, $\Psi$ are more general. This assertion can be justified by appealing to the WKB approximation; seeking solutions to ${\cal O}{\vec {\frak f}} = \lambda {\vec {\frak f}}$ of the form ${\vec {\frak f}} = \sum_{n=0}^\infty \check{\varepsilon}^n {\vec a}_n \exp(i\chi/\check{\varepsilon})$, where ${\vec a}_n$, $\chi$ are fields and $\check{\varepsilon}$ is the WKB expansion parameter, leads to a pair of eigenvalues that satisfy (\ref{lambda_prod}) with $l_\mu$ substituted by $\partial_\mu \chi/\check{\varepsilon}$. Those eigenvalues are identical to the eigenvalues of ${\cal O}_+$, ${\cal O}_-$ in the WKB approximation and, hence, (\ref{Lambda_factor}), (\ref{Lambda_pm}) hold. This approach is analogous to using the Euler-Heisenberg action to describe QED vacuum polarisation even when the electromagnetic invariants are not constant. Although the derivation of the Euler-Heisenberg action requires the electromagnetic fields to be constant (the potentials are linear in Minkowski coordinates), it is not uncommon to use the result in more general circumstances.

Although one can replace (\ref{O_pm}) with more complicated second-order linear operators and, via the WKB approximation, motivate more complicated expressions for $\Lambda_+$, $\Lambda_-$, the choice (\ref{Lambda_pm}) is perhaps the most natural. This conclusion is supported by the special case in which $\eta^{\sigma\lambda}\partial_\sigma \Psi \partial_\lambda \Psi \ll -1$ and $\partial_\mu \Phi \approx \partial_\mu \Psi$. In this particular situation
\begin{align}
\label{quadratic_action_two_comp_special}
\Lambda[\vec f] \approx \int d^2 x \sqrt{-\eta}\,\frac{1}{2}\big\{\eta^{\mu\nu}\partial_\mu \Phi \partial_\nu \Phi\,(\eta^{\sigma\tau}\partial_\sigma f_1 \partial_\tau f_1 + \eta^{\sigma\tau}\partial_\sigma f_2\partial_\tau f_2) + 4\eta^{\mu\nu}\partial_\mu \Phi\partial_\nu f_1\,\eta^{\sigma\tau}\partial_\sigma \Phi\partial_\tau f_2\big\}
\end{align}
follows from (\ref{quadratic_action_two_comp}). Introducing the new variables $\check{f}_1 = (f_1 + f_2)/\sqrt{2}$, $\check{f}_2 = (- f_1 + f_2)/\sqrt{2}$ immediately yields
\begin{equation}
\Lambda \approx \int d^2 x \sqrt{-\eta}\,\frac{1}{2}\big({\cal A}^{\mu\nu}_- \partial_\mu \check{f}_1\partial_\nu \check{f}_1 + {\cal A}^{\mu\nu}_+ \partial_\mu \check{f}_2\partial_\nu \check{f}_2\big)
\end{equation}
where ${\cal A}^{\mu\nu}_+$, ${\cal A}^{\mu\nu}_-$ are given by (\ref{calA_+_def}), (\ref{calA_-_def}) with the substitutions $\partial_\mu \Psi \rightarrow \partial_\mu \Phi$, $\eta^{\sigma\tau}\partial_\sigma \Psi \partial_\tau \Psi + 1 \rightarrow \eta^{\sigma\tau}\partial_\sigma \Phi \partial_\tau \Phi$ and the choice $s=-1$. The latter is required because $\eta^{\mu\nu}\partial_\mu \Phi\partial_\nu \Phi < 0$.
The functional measure satisfies ${\cal D}{\vec f} = {\cal D}{\vec{\check{f}}}$ because the transformation between ${\vec f}$ and ${\vec{\check{f}}}$ is a constant rotation, and we obtain
\begin{equation}
\label{Lambda_approx_factor}
\int {\cal D}{\vec{\check{f}}}\exp(i\Lambda[\vec f]) \approx \bigg\{\int {\cal D}\check{f}_1\exp(i\Lambda_+[\check{f}_1])\bigg\}\bigg\{\int {\cal D}\check{f}_2\exp(i\Lambda_-[\check{f}_2])\bigg\}
\end{equation} 
as required. Unfortunately, when considered as a metric, ${\cal A}^{\mu\nu}_+$ does not have a Lorentzian signature in this case and the classical states in this regime are not perturbatively stable. Hence, the validity of the effective action $\Gamma$ is questionable in this regime. Nevertheless, the above considerations suggest the use of (\ref{calA_+_def}), (\ref{calA_-_def}), (\ref{Lambda_factor}), (\ref{Lambda_pm}) in cases where ${\cal A}_+^{\mu\nu}$, ${\cal A}_-^{\mu\nu}$ both have Lorentzian signatures. Furthermore, the above special case fixes the sign of $s$; in fact, we will see in Section~\ref{subsec:dispersion_relations} that, in general, the quantum theory has physically unreasonable implications if $s$ is positive. It is convenient to delay substituting $s$ until Section~\ref{subsec:dispersion_relations}.

The metric signatures of ${\cal A}^+_{\mu\nu}$, ${\cal A}^-_{\mu\nu}$ can be deduced from the signs of the eigenvalues of the tensors $M_+^\mu{ }_\nu$, $M_-^\mu{ }_\nu$ given by
\begin{equation}
M_+^\mu{ }_\nu = \delta^\mu_\nu + B\,\eta^{\mu\omega}(X_\omega Y_\nu + Y_\omega X_\nu),\quad M_-^\mu{ }_\nu = \delta^\mu_\nu - B\,\eta^{\mu\omega}(X_\omega Y_\nu + Y_\omega X_\nu),
\end{equation}
where the timelike unit normalised covector fields $X_\mu$, $Y_\nu$ and scalar field $B$ (satisfying $B>1$) are
\begin{equation}
X_\mu = \frac{\partial_\mu \Phi}{\sqrt{-\eta^{\sigma\tau}\partial_\sigma \Phi\partial_\tau \Phi}},\quad Y_\mu = \frac{\partial_\mu \Psi}{\sqrt{-\eta^{\sigma\tau}\partial_\sigma \Psi\partial_\tau \Psi}},\qquad B = \sqrt{\frac{\eta^{\sigma\tau}\partial_\sigma \Psi\partial_\tau \Psi}{\eta^{\gamma\delta}\partial_\gamma \Psi\partial_\delta \Psi+1}}.
\end{equation}
The eigenvalues of $M_+^\mu{ }_\nu$, $M_-^\mu{ }_\nu$ are $\{1-B(\cosh\chi + 1)$, $1-B(\cosh\chi - 1)\}$ and $\{1+B(\cos\chi + 1)$, $1+B(\cosh\chi - 1)\}$, respectively, where $\cosh\chi=-\eta^{\mu\nu} X_\mu Y_\nu$. By inspection, the eigenvalues of $M_-^\mu{ }_\nu$ are always positive and, since $M_-^\mu{ }_\nu = \eta_{\nu\sigma}s{\cal A}_-^{\sigma\mu}/C$ where $C$ is a positive scalar field, it follows that $s{\cal A}_-^{\mu\nu}$ has the same signature as $\eta^{\mu\nu}$. However, one of the eigenvalues of $M_+^\mu{ }_\nu = \eta_{\nu\sigma}s{\cal A}_+^{\sigma\mu}/C$ is always negative and the sign of the remaining eigenvalue depends on the properties of the fields. The remaining eigenvalue must be negative for $s{\cal A}_+^{\mu\nu}$ to be Lorentzian, albeit of opposite signature to $\eta^{\mu\nu}$. Thus, $\cosh\chi > (1+B)/B$, which can be expressed as the condition
\begin{equation}
\label{Lorentzian_condition}
-\eta^{\mu\nu}\partial_\mu \Phi \partial_\nu \Psi > \sqrt{\eta^{\sigma\tau}\partial_\sigma \Phi \partial_\tau \Phi\,(\eta^{\mu\nu}\partial_\mu \Psi \partial_\nu \Psi + 1)} + \sqrt{\eta^{\sigma\tau}\partial_\sigma \Phi \partial_\tau \Phi\,\eta^{\mu\nu}\partial_\mu \Psi \partial_\nu \Psi}.
\end{equation}
Henceforth, we will only consider the regime in which (\ref{Lorentzian_condition}) is satisfied.

An explicit expression for the effective action $\Gamma$ is readily obtained by appealing to studies of the behaviour of the quantum vacuum in curved spacetimes. The required results emerge when (\ref{Lambda_factor}) is expressed in terms of a massless field theory on a dilatonic curved background. The pair of metrics $g^+_{\mu\nu}$, $g^-_{\mu\nu}$ and the pair of dilatons $\varphi^+$, $\varphi^-$ are
\begin{align}
\label{g_phi_def}
g^{\mu\nu}_+ = \frac{{\cal A}^{\mu\nu}_+}{\sqrt{A_+}},\qquad \varphi_+ = -\frac{1}{4}\ln(A_+),\qquad g^{\mu\nu}_- = \frac{{\cal A}^{\mu\nu}_-}{\sqrt{A_-}},\qquad \varphi_- = -\frac{1}{4}\ln(A_-)
\end{align}
where $A_+$, $A_-$ are the determinants of the tensors $A_+^{\,\,\nu}{ }_\mu = \eta_{\mu\sigma}{\cal A}^{\sigma\nu}_+$, $A_-^{\,\,\nu}{ }_\mu = \eta_{\mu\sigma}{\cal A}^{\sigma\nu}_-$, respectively. It follows that (\ref{Lambda_pm}) can be expressed as
\begin{align}
\Lambda_+[f] = \int d^2 x \sqrt{-g^+}\,\frac{1}{2} \exp(-2\varphi_+)g_+^{\mu\nu}\partial_\mu f \partial_\nu f,\quad
\Lambda_-[f] = \int d^2 x \sqrt{-g^-}\,\frac{1}{2} \exp(-2\varphi_-)g_-^{\mu\nu}\partial_\mu f \partial_\nu f
\end{align}
with $g^+$, $g^-$ the determinants of $g^+_{\mu\nu}$, $g^-_{\mu\nu}$, respectively, and the effective action (\ref{1-loop_action}) decomposes as
\begin{equation}
\label{1-loop_action_decomposed}
\Gamma = S - i\ln\bigg\{\int{\cal D}f\,\exp(i\Lambda_+[f])\bigg\} - i\ln\bigg\{\int{\cal D}f\,\exp(i\Lambda_-[f])\bigg\}.
\end{equation}

The effective action $W$ given by
\begin{align}
\label{2d_dilaton_gravity}
\exp(iW[g_{\mu\nu},\varphi]) = \int{\cal D}f\,\exp\bigg\{\frac{i}{2}\int d^2 x \sqrt{-g}\, \exp(-2\varphi)(\nabla f)^2\bigg\}
\end{align}
describes the coupling of a dilaton $\varphi$ to a Lorentzian metric $g_{\mu\nu}$, and their self-couplings, due to the vacuum fluctuations of a massless scalar field $f$. Its exact renormalised form is~\cite{kummer:1999, kummer:1999b}
\begin{align}
\label{kummer_eff_action}
W[g_{\mu\nu},\varphi] = &\frac{1}{24\pi}\int d^2 x \sqrt{-g} \bigg\{\frac{1}{4} R\Box^{-1}R + R(\psi + 2\varphi) - 3(\nabla\varphi)^2\Box^{-1}R - (\nabla\varphi)^2 - 4\nabla\varphi\cdot\nabla\psi - (\nabla\psi)^2\bigg\}\\
\notag
&- \frac{1}{4\pi}\ln\mu\int d^2 x \sqrt{-g} (\nabla\varphi)^2
\end{align}  
where $(\nabla f)^2 = \nabla f \cdot \nabla f$, $\nabla \varphi \cdot\nabla \psi = g^{\mu\nu} \nabla_\mu\varphi\nabla_\nu\psi$, $\Box=g^{\mu\nu}\nabla_\mu\nabla_\nu$, $\nabla_\mu$ is the Levi-Civita covariant derivative given by $g_{\mu\nu}$, and $R$ is the scalar curvature of $g_{\mu\nu}$. The conventions used for the Riemann tensor and Ricci tensor underpinning $R$ are given in Ref.~\cite{vassilevich:2003}. The scalar field $\psi$ that appears in (\ref{kummer_eff_action}) captures some of the freedom in the choice of the measure ${\cal D}f$. In particular, the quantity $\int{\cal D} f \exp(i\langle f, f\rangle)$ is chosen to be a field-independent constant, where the inner product $\langle\cdot,\cdot\rangle$ is given by $\langle a, b\rangle = \int d^2 x \sqrt{-g} \exp(-2\psi)\, a^*\,b$. The constant $\mu$ in (\ref{kummer_eff_action}) emerges from the zeta-function regularisation technique used to derive (\ref{kummer_eff_action}) and, in general, must be fixed using additional information such as experimental data.

The result of each functional integral in (\ref{1-loop_action_decomposed}) follows immediately from (\ref{kummer_eff_action}) using the respective substitutions $g_{\mu\nu} = g^+_{\mu\nu}$, $\varphi = \varphi^+$, $\mu=\mu^+$ and $g_{\mu\nu} = g^-_{\mu\nu}$, $\varphi = \varphi^-$, $\mu=\mu^-$. The inner product $\int d^2 x \sqrt{-\eta}\,a^*\,b$ induced from the background Minkowski metric $\eta_{\mu\nu}$ is natural in the present context; thus, since $g^+ = g^- = \eta$ follows from (\ref{g_phi_def}), we set $\psi=0$ in $\langle\cdot,\cdot\rangle$. In summary, an effective theory describing the self-interaction of a laser-driven plasma due to quantum fluctuations is
\begin{equation}
\label{GammaAction}
\Gamma[\Phi,\Psi] = S[\Phi,\Psi] + w[g^+_{\mu\nu},\varphi_+,\mu_+] + w[g^-_{\mu\nu},\varphi_-,\mu_-]
\end{equation}
where
\begin{equation}
\label{waction}
w[g_{\mu\nu},\varphi,\mu] = \frac{1}{24\pi}\int d^2 x \sqrt{-g} \bigg\{\frac{1}{4} R\Box^{-1}R + 2R\varphi - 3(\nabla\varphi)^2\Box^{-1}R - (\nabla\varphi)^2\bigg\} - \frac{1}{4\pi}\ln\mu\int d^2 x \sqrt{-g} (\nabla\varphi)^2
\end{equation}
and $g^+_{\mu\nu}$, $g^-_{\mu\nu}$, $\varphi^+$, $\varphi^-$ depend on $\Phi$, $\Psi$ according to (\ref{calA_+_def}), (\ref{calA_-_def}), (\ref{g_phi_def}). The fields must satisfy $\eta^{\mu\nu}\partial_\mu\Phi\partial_\nu\Phi < 0$, $\eta^{\mu\nu}\partial_\mu\Psi\partial_\nu\Psi < -1$, and the condition (\ref{Lorentzian_condition}).
\section{Field equations for $\Phi$ and $\Psi$}
\label{sec:field_equations}
Stationary variations of the action (\ref{GammaAction}) with respect to $\Phi$, $\Psi$ lead to field equations describing a laser-driven plasma that include the backreaction of the quantum fluctuations. The field equations arising from the $\Phi$ variation and $\Psi$ variation can be expressed as
\begin{equation}
\label{full_field_equations}
\nabla_\sigma^{(\eta)}(\varepsilon\beta^\sigma + {\cal B}_+^\sigma + {\cal B}_-^\sigma) = 0,\qquad\nabla_\sigma^{(\eta)}(\alpha\zeta^\sigma + {\cal C}_+^\sigma + {\cal C}_-^\sigma) = 0,
\end{equation}
respectively, where\footnote{The scalar field $\alpha$ is unrelated to the fine-structure constant.}
\begin{equation}
\label{alpha_defs}
\alpha=\eta^{\sigma\tau}\partial_{\sigma}\Phi\partial_\tau\Phi,\qquad\beta^{\mu}=\eta^{\mu\nu}\partial_\nu\Phi,\qquad\varepsilon=\eta^{\sigma\tau}\partial_{\sigma}\Psi\partial_{\tau}\Psi+1,\qquad\zeta^{\mu}=\eta^{\mu\nu}\partial_\nu\Psi,
\end{equation}
with
\begin{align}
\label{calB_def}
&{\cal B}_+^\sigma = \frac{1}{\sqrt{-\eta}}\frac{\delta w_+}{\delta{\cal A}^{\mu\nu}_+}\bigg(-s\eta^{\mu\nu}\sqrt{\frac{\varepsilon}{\alpha}}\beta^\sigma + 2s\zeta^{\mu} \eta^{\nu\sigma}\bigg),\qquad {\cal B}_-^\sigma = \frac{1}{\sqrt{-\eta}}\frac{\delta w_-}{\delta{\cal A}^{\mu\nu}_-}\bigg(-s\eta^{\mu\nu}\sqrt{\frac{\varepsilon}{\alpha}}\beta^\sigma - 2s\zeta^{\mu} \eta^{\nu\sigma}\bigg),\\
\label{calC_def}
&{\cal C}_+^\sigma = \frac{1}{\sqrt{-\eta}}\frac{\delta w_+}{\delta{\cal A}^{\mu\nu}_+}\bigg(-s\eta^{\mu\nu}\sqrt{\frac{\alpha}{\varepsilon}}\zeta^\sigma + 2s\beta^{\mu} \eta^{\nu\sigma}\bigg),\qquad {\cal C}_-^\sigma = \frac{1}{\sqrt{-\eta}}\frac{\delta w_-}{\delta{\cal A}^{\mu\nu}_-}\bigg(-s\eta^{\mu\nu}\sqrt{\frac{\alpha}{\varepsilon}}\zeta^\sigma - 2s\beta^{\mu} \eta^{\nu\sigma}\bigg),
\end{align}
where $w_+ = w[g^+_{\mu\nu},\varphi_+,\mu_+]$, $w_- = w[g^-_{\mu\nu},\varphi_-,\mu_-]$. The minus sign in front of each square root inside the parentheses in (\ref{calB_def}), (\ref{calC_def}) arises because $\alpha, \varepsilon < 0$.

The structure of (\ref{calB_def}), (\ref{calC_def}) follows because the effective metrics $g^+_{\mu\nu}$, $g^-_{\mu\nu}$ and dilatons $\varphi_+$, $\varphi_-$ can be expressed solely in terms of ${\cal A}^{\mu\nu}_+$, ${\cal A}^{\mu\nu}_-$, respectively. The details follow using
\begin{equation}
\delta w_+ = \int d^2 x \frac{\delta w_+}{\delta{\cal A}^{\mu\nu}_+}\delta{\cal A}^{\mu\nu}_+,\qquad \delta w_- = \int d^2 x \frac{\delta w_-}{\delta{\cal A}^{\mu\nu}_-}\delta{\cal A}^{\mu\nu}_-,
\end{equation}
with
\begin{align}
&\delta{\cal A}^{\mu\nu}_+ = \bigg(-s\eta^{\mu\nu}\sqrt{\frac{\varepsilon}{\alpha}}\beta^\sigma + 2s\zeta^{(\mu} \eta^{\nu)\sigma}\bigg)\partial_\sigma\delta\Phi + \bigg(-s\eta^{\mu\nu}\sqrt{\frac{\alpha}{\varepsilon}}\zeta^\sigma + 2s\beta^{(\mu} \eta^{\nu)\sigma}\bigg)\partial_\sigma\delta\Psi,\\
&\delta{\cal A}^{\mu\nu}_- = \bigg(-s\eta^{\mu\nu}\sqrt{\frac{\varepsilon}{\alpha}}\beta^\sigma - 2s\zeta^{(\mu} \eta^{\nu)\sigma}\bigg)\partial_\sigma\delta\Phi + \bigg(-s\eta^{\mu\nu}\sqrt{\frac{\alpha}{\varepsilon}}\zeta^\sigma - 2s\beta^{(\mu} \eta^{\nu)\sigma}\bigg)\partial_\sigma\delta\Psi,
\end{align}
which emerge from (\ref{calA_+_def}), (\ref{calA_-_def}). The parentheses enclosing indices denote symmetrisation with the standard weighting; e.g. $2\beta^{(\mu} \eta^{\nu)\sigma} = \beta^{\mu} \eta^{\nu\sigma} + \beta^{\nu} \eta^{\mu\sigma}$. As usual, the variations $\delta\Phi$, $\delta\Psi$ are chosen to have compact support; thus, no boundary terms arise during the derivation of the field equations.

The remainder of this section is focussed on determining the functional derivatives of $w_+$, $w_-$ with respect to ${\cal A}^{\mu\nu}_+$, ${\cal A}^{\mu\nu}_-$, respectively. To achieve this goal, it is fruitful to briefly return to the most natural variables for expressing $w_+$, $w_-$; the effective metrics and dilatons.
\subsection{Variations of $w$ with respect to $g^{\mu\nu}$ and $\varphi$}
\label{sec:variations_metric_dilaton}
Since $w_+$ (or $w_-$) is simply $w$ evaluated at particular values of its arguments, we can capture the variations of $w_+$ with respect to $g^{\mu\nu}_+$, $\varphi_+$ (or $g^{\mu\nu}_-$, $\varphi_-$) by appealing solely to the variations of $w$ with respect to $g^{\mu\nu}$, $\varphi$.
 
Taking care of the inverse D'Alembertian operator $\Box^{-1}$ using the techniques given in Ref.~\cite{conroy:2014}, we find that the functional derivatives of $w$ with respect to $g^{\mu\nu}$, $\varphi$ are
\begin{align}
\label{fd_w_metric}
\frac{48\pi}{\sqrt{-g}}\frac{\delta w}{\delta g^{\mu\nu}} &=
-\nabla_{\mu}\nabla_{\nu}\Box^{-1}R+g_{\mu\nu}R-\frac{1}{4}g_{\mu\nu}\nabla^{\sigma}\Box^{-1}R\nabla_{\sigma}\Box^{-1}R+\frac{1}{2}\nabla_{\nu}\Box^{-1}R\nabla_{\mu}\Box^{-1}R
\nonumber\\&
+3g_{\mu\nu}(\nabla\varphi)^{2}\Box^{-1}R+3g_{\mu\nu}\nabla^{\sigma}\Box^{-1}(\nabla\varphi)^{2}\nabla_{\sigma}\Box^{-1}R
-6\nabla_{(\nu}\Box^{-1}(\nabla\varphi)^{2}\nabla_{\mu)}\Box^{-1}R
\nonumber\\
&-6(\nabla_{\mu}\varphi\nabla_{\nu}\varphi)\Box^{-1}R
+6\nabla_{\mu}\nabla_{\nu}\Box^{-1}(\nabla\varphi)^{2}-4\nabla_{\mu}\nabla_{\nu}\varphi+4g_{\mu\nu}\Box\varphi
\nonumber\\&+g_{\mu\nu}\left(-5+6\ln\mu\right)(\nabla\varphi)^{2}-2(\nabla_{\mu}\varphi\nabla_{\nu}\varphi)(1+6\ln\mu),\\
\label{fd_w_dilaton}
\frac{12\pi}{\sqrt{-g}}\frac{\delta w}{\delta\varphi}&= 3\nabla_{\mu}\Box^{-1}R\nabla^{\mu}\varphi+\left(3\Box^{-1}R+1+6\ln\mu\right)\Box\varphi + R.
\end{align}
For convenience, indices have been lowered (or raised) using the metric tensor $g_{\mu\nu}$ (or its inverse $g^{\mu\nu}$) in (\ref{fd_w_metric}), (\ref{fd_w_dilaton}).
\subsection{Variation of $w$ with respect to ${\cal A}^{\mu\nu}$}
The appropriate combination of (\ref{fd_w_metric}), (\ref{fd_w_dilaton}) that appears in the field equations (\ref{full_field_equations}) emerges upon introducing the variable ${\cal A}^{\mu\nu} = e^{-2\varphi}g^{\mu\nu}$, where $4\varphi = -\ln(A)$ with $A$ the determinant of $A^\nu{ }_\mu = \eta_{\mu\sigma}{\cal A}^{\sigma\nu}$. Hence, $\delta g^{\mu\nu} = e^{2\varphi}\delta{\cal A}^{\mu\nu} + 2g^{\mu\nu}\delta\varphi$ and $4\delta\varphi = -e^{2\varphi}g_{\mu\nu}\delta{\cal A}^{\mu\nu}$. It follows
\begin{equation}
\delta w = \int d^2 x \bigg(\frac{\delta w}{\delta g^{\mu\nu}}\delta g^{\mu\nu} + \frac{\delta w}{\delta\varphi}\delta\varphi\bigg) = \int d^2 x \bigg[\frac{\delta w}{\delta g^{\mu\nu}}- \frac{1}{4}\bigg(2\frac{\delta w}{\delta g^{\sigma\tau}}g^{\sigma\tau} + \frac{\delta w}{\delta\varphi}\bigg)g_{\mu\nu}\bigg]e^{2\varphi}\delta{\cal A}^{\mu\nu}
\end{equation}
and we obtain
\begin{equation}
\label{fd_w_A_expressed}
\frac{1}{\sqrt{-\eta}}\frac{\delta w}{\delta{\cal A}^{\mu\nu}} = \frac{1}{\sqrt{-g}}\bigg[\frac{\delta w}{\delta g^{\mu\nu}}- \frac{1}{4}\bigg(2\frac{\delta w}{\delta g^{\sigma\tau}}g^{\sigma\tau} + \frac{\delta w}{\delta\varphi}\bigg)g_{\mu\nu}\bigg]e^{2\varphi}.
\end{equation}
The equality of the determinants of $\eta_{\mu\nu}$ and $g_{\mu\nu}$ has been used to express (\ref{fd_w_A_expressed}) in a convenient form.

Hence, the field equations (\ref{full_field_equations}) for $\Phi$, $\Psi$ are specified by substituting the functional derivatives found in (\ref{calB_def}), (\ref{calC_def}) with
\begin{align}
\label{fd_w+_A+}
&\frac{1}{\sqrt{-\eta}}\frac{\delta w_+}{\delta{\cal A}_+^{\mu\nu}} = \frac{1}{\sqrt{-g^+}}\bigg[\frac{\delta w_+}{\delta g_+^{\mu\nu}}- \frac{1}{4}\bigg(2\frac{\delta w_+}{\delta g_+^{\sigma\tau}}g_+^{\sigma\tau} + \frac{\delta w_+}{\delta\varphi_+}\bigg)g^+_{\mu\nu}\bigg]e^{2\varphi_+},\\
\label{fd_w-_A-}
&\frac{1}{\sqrt{-\eta}}\frac{\delta w_-}{\delta{\cal A}_-^{\mu\nu}} = \frac{1}{\sqrt{-g^-}}\bigg[\frac{\delta w_-}{\delta g_-^{\mu\nu}}- \frac{1}{4}\bigg(2\frac{\delta w_-}{\delta g_-^{\sigma\tau}}g_-^{\sigma\tau} + \frac{\delta w_-}{\delta\varphi_-}\bigg)g^-_{\mu\nu}\bigg]e^{2\varphi_-}.
\end{align}
The right-hand side of (\ref{fd_w+_A+}) (or (\ref{fd_w-_A-})) is given by substituting $g^+_{\mu\nu}$, $\varphi_+$, $\mu_+$ (or $g^-_{\mu\nu}$, $\varphi_-$, $\mu_-$) into (\ref{fd_w_metric}), (\ref{fd_w_dilaton}).
\section{Linearised field equations}
\label{sec:linearised_field_equations}
The simplest exact solutions to (\ref{full_field_equations}) describe a uniform, monochromatic, laser beam propagating through a uniform plasma. In this case, $\beta^\mu$, $\zeta^\nu$ are covariantly constant with respect to the Levi-Civita connection $\nabla^{(\eta)}$ of the flat spacetime metric $\eta_{\mu\nu}$. Using (\ref{alpha_defs}), it is clear that the classical terms in (\ref{full_field_equations}) immediately vanish. The dilatons $\varphi_+$, $\varphi_-$ and effective metrics $g^+_{\mu\nu}$, $g^-_{\mu\nu}$ are constructed solely from tensors that are covariantly constant with respect to $\nabla^{(\eta)}$; thus, they are also covariantly constant with respect to $\nabla^{(\eta)}$. It follows that the components of the effective metrics are constant in a Minkowski coordinate system adapted to $\eta_{\mu\nu}$; thus, their Christoffel symbols are zero. In addition to the fact that the dilatons are constant, we conclude that the curvatures of the effective metrics are zero. Inspection of (\ref{fd_w_metric}), (\ref{fd_w_dilaton}) shows that the quantum corrections to the classical field equations are zero as required.

We will now uncover the impact of the quantum backreaction on perturbations to the exact solutions describing a uniform, monochromatic, laser beam propagating through a uniform plasma. Throughout the following, we will use a `bar' to denote fields and operators associated with the unperturbed exact solutions. For simplicity, we will use Minkowski coordinates adapted to $\eta_{\mu\nu}$; thus, for the reasons given above, all components of `bar' tensors are constant.

Introducing the substitutions 
\begin{equation}
g^{\mu\nu} = \bar{g}^{\mu\nu} + g^{\mu\nu}_{(1)},\qquad \varphi = \bar{\varphi} + \varphi_{(1)}
\end{equation}
in (\ref{fd_w_metric}), (\ref{fd_w_dilaton}) gives
\begin{align}
\label{lin_fd_w_g}
&\frac{48\pi}{\sqrt{-g}}\frac{\delta w}{\delta g^{\mu\nu}} =
-\partial_{\mu}\partial_{\nu}\bar\Box^{-1}R_{(1)} + \bar{g}_{\mu\nu}R_{(1)} - 4\partial_\mu\partial_\nu\varphi_{(1)} + 4\bar{g}_{\mu\nu}\bar\Box\varphi_{(1)},\\
\label{lin_fd_w_phi}
&\frac{12\pi}{\sqrt{-g}}\frac{\delta w}{\delta\varphi} = R_{(1)} + (1+6\ln\mu)\bar\Box\varphi_{(1)} 
\end{align}
to first order in the perturbations $g^{\mu\nu}_{(1)}$, $\varphi_{(1)}$. The scalar curvature perturbation $R_{(1)}$ is
\begin{equation}
\label{R1_def}
R_{(1)} = - \partial_\mu\partial_\nu g^{\mu\nu}_{(1)} + \bar{g}_{\mu\nu}\bar\Box g^{\mu\nu}_{(1)},
\end{equation}
and $\bar\Box = \bar{g}^{\mu\nu}\partial_\mu\partial_\nu$. 

Hence, (\ref{full_field_equations}), (\ref{calB_def}), (\ref{calC_def}), (\ref{fd_w+_A+}), (\ref{fd_w-_A-}), (\ref{lin_fd_w_g}), (\ref{lin_fd_w_phi}) together give
\begin{align}
\nonumber
&0=\bar\varepsilon\Box_{(\eta)}\Phi_{(1)} + 2\bar\beta^\mu\bar\zeta^\nu\partial_\mu\partial_\nu\Psi_{(1)}\\
\nonumber
&\quad - \frac{s}{96\pi}\sqrt{\frac{\bar\varepsilon}{\bar\alpha}}\bar\beta^\mu\partial_\mu\big\{e^{2\bar\varphi_+}\big[-2(\Box_{(\eta)}/\bar{\Box}^+) R^+_{(1)}-c^+ R^+_{(1)} - 8\Box_{(\eta)}\varphi^+_{(1)} + 2c^+(1-6\ln\mu_+)\bar\Box^+\varphi^+_{(1)}\big]\\
\nonumber
&\qquad\qquad\qquad\quad\,\,\,+ e^{2\bar\varphi_-}\big[ -2(\Box_{(\eta)}/\bar{\Box}^-) R^-_{(1)} - c^- R^-_{(1)} - 8\Box_{(\eta)}\varphi^-_{(1)} + 2c^-(1-6\ln\mu_-)\bar\Box^-\varphi^-_{(1)}
\big]\big\}\\
\nonumber
&\quad +\frac{s}{48\pi}\bar\zeta^\mu\big\{e^{2\bar\varphi_+}\big[-2(\Box_{(\eta)}/\bar\Box^+)\partial_\mu R^+_{(1)} - \bar{g}^+_{\mu\nu}\partial^\nu R^+_{(1)} - 8\partial_\mu\Box_{(\eta)}\varphi^+_{(1)} + 2(1-6\ln\mu_+)\bar{g}^+_{\mu\nu}\partial^\nu\varphi^+_{(1)}\big]\\
\label{Phi_lin_eq}
&\qquad\quad\quad\,\,\, - e^{2\bar\varphi_-}\big[-2(\Box_{(\eta)}/\bar\Box^-)\partial_\mu R^-_{(1)} - \bar{g}^-_{\mu\nu}\partial^\nu R^-_{(1)} - 8\partial_\mu\Box_{(\eta)}\varphi^-_{(1)} + 2(1-6\ln\mu_-)\bar{g}^-_{\mu\nu}\partial^\nu\varphi^-_{(1)}\big]\big\},
\end{align}
and
\begin{align}
\nonumber
&0=\bar\alpha\Box_{(\eta)}\Psi_{(1)} + 2\bar\beta^\mu\bar\zeta^\nu\partial_\mu\partial_\nu\Phi_{(1)}\\
\nonumber
&\quad - \frac{s}{96\pi}\sqrt{\frac{\bar\alpha}{\bar\varepsilon}}\bar\zeta^\mu\partial_\mu\big\{e^{2\bar\varphi_+}\big[-2(\Box_{(\eta)}/\bar{\Box}^+) R^+_{(1)}-c^+ R^+_{(1)} - 8\Box_{(\eta)}\varphi^+_{(1)} + 2c^+(1-6\ln\mu_+)\bar\Box^+\varphi^+_{(1)}\big]\\
\nonumber
&\qquad\qquad\qquad\quad\,\,\,+ e^{2\bar\varphi_-}\big[ -2(\Box_{(\eta)}/\bar{\Box}^-) R^-_{(1)} - c^- R^-_{(1)} - 8\Box_{(\eta)}\varphi^-_{(1)} + 2c^-(1-6\ln\mu_-)\bar\Box^-\varphi^-_{(1)}
\big]\big\}\\
\nonumber
&\quad +\frac{s}{48\pi}\bar\beta^\mu\big\{e^{2\bar\varphi_+}\big[-2(\Box_{(\eta)}/\bar\Box^+)\partial_\mu R^+_{(1)} - \bar{g}^+_{\mu\nu}\partial^\nu R^+_{(1)} - 8\partial_\mu\Box_{(\eta)}\varphi^+_{(1)} + 2(1-6\ln\mu_+)\bar{g}^+_{\mu\nu}\partial^\nu\varphi^+_{(1)}\big]\\
\label{Psi_lin_eq}
&\qquad\quad\quad\,\,\, - e^{2\bar\varphi_-}\big[-2(\Box_{(\eta)}/\bar\Box^-)\partial_\mu R^-_{(1)} - \bar{g}^-_{\mu\nu}\partial^\nu R^-_{(1)} - 8\partial_\mu\Box_{(\eta)}\varphi^-_{(1)} + 2(1-6\ln\mu_-)\bar{g}^-_{\mu\nu}\partial^\nu\varphi^-_{(1)}\big]\big\},
\end{align}
where $\partial^\mu = \eta^{\mu\nu}\partial_\nu$, $\Box_{(\eta)} = \eta^{\mu\nu}\partial_\mu\partial_\nu$, $c^+=\bar{g}^{\mu\nu}_+\eta_{\mu\nu}$,  $c^-=\bar{g}^{\mu\nu}_-\eta_{\mu\nu}$, and
\begin{equation}
\label{R1+-_def}
R^+_{(1)} = - \partial_\mu\partial_\nu g^{\mu\nu}_{+(1)} + \bar{g}^+_{\mu\nu}\bar\Box^+ g^{\mu\nu}_{+(1)},\qquad R^-_{(1)} = - \partial_\mu\partial_\nu g^{\mu\nu}_{-(1)} + \bar{g}^-_{\mu\nu}\bar\Box^- g^{\mu\nu}_{-(1)}.
\end{equation}
The perturbations to the effective metrics and dilatons are given in terms of $\Phi_{(1)}$, $\Psi_{(1)}$ by
\begin{align}
&g^{\mu\nu}_{+(1)} = 2\varphi^+_{(1)} \bar{g}^{\mu\nu}_{+} + e^{2\bar{\varphi}_+} {\cal A}^{\mu\nu}_{+(1)},\qquad g^{\mu\nu}_{-(1)} = 2\varphi^-_{(1)} \bar{g}^{\mu\nu}_{-} + e^{2\bar{\varphi}_-} {\cal A}^{\mu\nu}_{-(1)},\\
&4\varphi^+_{(1)} = -e^{2\bar\varphi_+} \bar{g}^+_{\mu\nu}{\cal A}^{\mu\nu}_{+(1)},\qquad 4\varphi^-_{(1)} = -e^{2\bar\varphi_-} \bar{g}^-_{\mu\nu}{\cal A}^{\mu\nu}_{-(1)},
\end{align}
with
\begin{align}
\label{calA1+_def}
&{\cal A}^{\mu\nu}_{+(1)} = \bigg(-s\eta^{\mu\nu}\sqrt{\frac{\bar\varepsilon}{\bar\alpha}}\bar\beta^\sigma + 2s\bar\zeta^{(\mu} \eta^{\nu)\sigma}\bigg)\partial_\sigma\Phi_{(1)} + \bigg(-s\eta^{\mu\nu}\sqrt{\frac{\bar\alpha}{\bar\varepsilon}}\bar\zeta^\sigma + 2s\bar\beta^{(\mu} \eta^{\nu)\sigma}\bigg)\partial_\sigma\Psi_{(1)},\\
\label{calA1-_def}
&{\cal A}^{\mu\nu}_{-(1)} = \bigg(-s\eta^{\mu\nu}\sqrt{\frac{\bar\varepsilon}{\bar\alpha}}\bar\beta^\sigma - 2s\bar\zeta^{(\mu} \eta^{\nu)\sigma}\bigg)\partial_\sigma\Phi_{(1)} + \bigg(-s\eta^{\mu\nu}\sqrt{\frac{\bar\alpha}{\bar\varepsilon}}\bar\zeta^\sigma - 2s\bar\beta^{(\mu} \eta^{\nu)\sigma}\bigg)\partial_\sigma\Psi_{(1)}.
\end{align}
\subsection{Plane-wave perturbations and their dispersion relations}
\label{subsec:dispersion_relations}
Inspection of (\ref{Phi_lin_eq}), (\ref{Psi_lin_eq}) shows that the classical behaviour of the perturbations $\Phi_{(1)}$, $\Psi_{(1)}$ is determined by the linear equations
\begin{align}
\label{classical_linearised}
\bar\varepsilon\Box_{(\eta)}\Phi_{(1)} + 2\bar\beta^\mu\bar\zeta^\nu\partial_\mu\partial_\nu\Psi_{(1)} = 0,\qquad\bar\alpha\Box_{(\eta)}\Psi_{(1)} + 2\bar\beta^\mu\bar\zeta^\nu\partial_\mu\partial_\nu\Phi_{(1)}=0.
\end{align}
Using the plane-wave ans\"atze $\Phi_{(1)} \propto \exp(ik x)$,  $\Psi_{(1)} \propto \exp(ik x)$ in (\ref{classical_linearised}), where $k x \equiv k_\mu x^\mu$, leads to the dispersion relation
$\bar{\cal A}_+^{\mu\nu}k_\mu k_\nu\,\bar{\cal A}_-^{\sigma\omega}k_\sigma k_\omega = 0$ where $\bar{\cal A}_+^{\mu\nu} = s\sqrt{\bar\alpha\bar\varepsilon}\,\eta^{\mu\nu} + 2s\bar\beta^{(\mu}\bar\zeta^{\nu)}$, $\bar{\cal A}_-^{\mu\nu} = s\sqrt{\bar\alpha\bar\varepsilon}\,\eta^{\mu\nu} - 2s\bar\beta^{(\mu}\bar\zeta^{\nu)}$. Furthermore, the presence of $1/\bar\Box^+$ within the quantum corrections in (\ref{Phi_lin_eq}), (\ref{Psi_lin_eq}) suggests that the terms denoted ``$\dots$'' inside the equations
\begin{align}
\nonumber
&0=\bar\varepsilon\Box_{(\eta)}\Phi_{(1)} + 2\bar\beta^\mu\bar\zeta^\nu\partial_\mu\partial_\nu\Psi_{(1)}\\
\label{Phi_lin_eq_dots_plus}
&\quad - \frac{s}{96\pi}\sqrt{\frac{\bar\varepsilon}{\bar\alpha}}\bar\beta^\mu\partial_\mu\big\{e^{2\bar\varphi_+}\big[-2(\Box_{(\eta)}/\bar{\Box}^+) R^+_{(1)}+\dots\big]\big\} +\frac{s}{48\pi}\bar\zeta^\mu\big\{e^{2\bar\varphi_+}\big[-2(\Box_{(\eta)}/\bar\Box^+)\partial_\mu R^+_{(1)} + \dots\big]\big\},\\
\nonumber
&0=\bar\alpha\Box_{(\eta)}\Psi_{(1)} + 2\bar\beta^\mu\bar\zeta^\nu\partial_\mu\partial_\nu\Phi_{(1)}\\
\label{Psi_lin_eq_dots_plus}
&\quad - \frac{s}{96\pi}\sqrt{\frac{\bar\alpha}{\bar\varepsilon}}\bar\zeta^\mu\partial_\mu\big\{e^{2\bar\varphi_+}\big[-2(\Box_{(\eta)}/\bar{\Box}^+) R^+_{(1)}+\dots\big]\big\} +\frac{s}{48\pi}\bar\beta^\mu\big\{e^{2\bar\varphi_+}\big[-2(\Box_{(\eta)}/\bar\Box^+)\partial_\mu R^+_{(1)} + \dots\big]\big\}
\end{align}
are negligible close to the classical solution satisfying $\bar{\cal A}_+^{\mu\nu}k_\mu k_\nu = 0$. Likewise, the presence of $1/\bar\Box^-$ within the quantum corrections in (\ref{Phi_lin_eq}), (\ref{Psi_lin_eq}) suggests that the terms denoted ``$\dots$'' inside
\begin{align}
\nonumber
&0=\bar\varepsilon\Box_{(\eta)}\Phi_{(1)} + 2\bar\beta^\mu\bar\zeta^\nu\partial_\mu\partial_\nu\Psi_{(1)}\\
\label{Phi_lin_eq_dots_minus}
&\quad - \frac{s}{96\pi}\sqrt{\frac{\bar\varepsilon}{\bar\alpha}}\bar\beta^\mu\partial_\mu\big\{e^{2\bar\varphi_-}\big[-2(\Box_{(\eta)}/\bar{\Box}^-) R^-_{(1)}+\dots\big]\big\} +\frac{s}{48\pi}\bar\zeta^\mu\big\{-e^{2\bar\varphi_-}\big[-2(\Box_{(\eta)}/\bar\Box^-)\partial_\mu R^-_{(1)} + \dots\big]\big\},\\
\nonumber
&0=\bar\alpha\Box_{(\eta)}\Psi_{(1)} + 2\bar\beta^\mu\bar\zeta^\nu\partial_\mu\partial_\nu\Phi_{(1)}\\
\label{Psi_lin_eq_dots_minus}
&\quad - \frac{s}{96\pi}\sqrt{\frac{\bar\alpha}{\bar\varepsilon}}\bar\zeta^\mu\partial_\mu\big\{e^{2\bar\varphi_-}\big[-2(\Box_{(\eta)}/\bar{\Box}^-) R^-_{(1)}+\dots\big]\big\} +\frac{s}{48\pi}\bar\beta^\mu\big\{-e^{2\bar\varphi_-}\big[-2(\Box_{(\eta)}/\bar\Box^-)\partial_\mu R^-_{(1)} + \dots\big]\big\}
\end{align}
are negligible close to the classical solution satisfying $\bar{\cal A}_-^{\mu\nu}k_\mu k_\nu = 0$. Note that the unknown constants $\mu_+$, $\mu_-$ in (\ref{Phi_lin_eq}), (\ref{Psi_lin_eq}) do not contribute in this regime.

Focussing on (\ref{Phi_lin_eq_dots_plus}), (\ref{Psi_lin_eq_dots_plus}), the above considerations suggest a perturbative analysis of
\begin{align}
\label{Phi_lin_eq_nearly}
&0=\bar g^{\mu\nu}_+ k_\mu k_\nu(\bar\varepsilon\, k\cdot k\,\Phi_{(1)} + 2\bar\beta k\,\bar\zeta k\,\Psi_{(1)}) + s\epsilon^2\frac{e^{2\bar\varphi_+}}{48\pi}\bigg(-\sqrt{\frac{\bar\varepsilon}{\bar\alpha}}i\bar\beta k + 2i\bar\zeta k \bigg)k\cdot k\,R^+_{(1)} + {\cal O}(\epsilon^3),\\
\label{Psi_lin_eq_nearly}
&0=\bar g^{\mu\nu}_+ k_\mu k_\nu(\bar\alpha\, k\cdot k\Psi_{(1)} + 2\bar\beta k\,\bar\zeta k\,\Phi_{(1)}) + s\epsilon^2\frac{e^{2\bar\varphi_+}}{48\pi}\bigg(-\sqrt{\frac{\bar\alpha}{\bar\varepsilon}}i\bar\zeta k + 2i\bar\beta k\bigg) k\cdot k\,R^+_{(1)} + {\cal O}(\epsilon^3),
\end{align}
where $\bar g^{\mu\nu}_+ k_\mu k_\nu = {\cal O}(\epsilon)$ is assumed and $k\cdot k \equiv \eta^{\mu\nu} k_\mu k_\nu$, $\bar\beta k \equiv \bar\beta^\mu k_\mu$, $\bar\zeta k \equiv \bar\zeta^\mu k_\mu$. The perturbation parameter $\epsilon$ has been introduced for clarity of exposition, and the $\epsilon$-orders of the terms have been allocated {\it a posteriori} so that the working is self-consistent. The parameter $\epsilon$ is merely a device for capturing perturbative orders, and can be set to unity at the end of the calculation. Thus, $R^+_{(1)} = k_\mu k_\nu g^{\mu\nu}_{+(1)} + {\cal O}(\epsilon)$ follows from (\ref{R1+-_def}), and so 
\begin{align}
\label{Phi_lin_eq_simp}
&\bar{g}^{\mu\nu}_+ k_\mu k_\nu (\bar\varepsilon k\cdot k\,\Phi_{(1)} + 2\bar\beta k\,\bar\zeta k\,\Psi_{(1)}) - \epsilon^2\frac{1}{48\pi}a(a\Phi_{(1)} + b\Psi_{(1)}) = {\cal O}(\epsilon^3),\\
\label{Psi_lin_eq_simp}
&\bar{g}^{\mu\nu}_+ k_\mu k_\nu (\bar\alpha k\cdot k\,\Psi_{(1)} + 2\bar\beta k\,\bar\zeta k\,\Phi_{(1)}) - \epsilon^2\frac{1}{48\pi}b(a\Phi_{(1)} + b\Psi_{(1)}) = {\cal O}(\epsilon^3)
\end{align}
emerge from (\ref{Phi_lin_eq_nearly}), (\ref{Psi_lin_eq_nearly}) using (\ref{calA1+_def}), where the quantities $a$, $b$ are
\begin{equation}
\label{ab_def}
a = \bigg(-\sqrt{\frac{\bar\varepsilon}{\bar\alpha}}\,\bar\beta k + 2\bar\zeta k\bigg)k\cdot k\, e^{2\bar\varphi_+},\qquad
b = \bigg(-\sqrt{\frac{\bar\alpha}{\bar\varepsilon}}\,\bar\zeta k + 2\bar\beta k\bigg)k\cdot k \,e^{2\bar\varphi_+}
\end{equation}
and $s^2=1$ has been used to simplify (\ref{Phi_lin_eq_simp}), (\ref{Psi_lin_eq_simp}), (\ref{ab_def}).

Equations (\ref{Phi_lin_eq_simp}), (\ref{Psi_lin_eq_simp}) together form a homogeneous linear system for $\Phi_{(1)}$, $\Psi_{(1)}$. For a non-zero solution to exist, the condition
\begin{equation}
\label{det_coeffs}
\bigg|\bar g^{\mu\nu}_+ k_\mu k_\nu
\begin{pmatrix}
\bar\varepsilon\, k\cdot k & 2\bar\beta k\,\bar\zeta k \\
2\bar\beta k\,\bar\zeta k & \bar\alpha\, k\cdot k\\
\end{pmatrix}
- \epsilon^2\frac{1}{48\pi}
\begin{pmatrix}
a^2 & ab \\
ab & b^2
\end{pmatrix}
\bigg| = {\cal O}(\epsilon^6)
\end{equation}
on the matrix determinant of the coefficients of the linear system must be satisfied. Equation (\ref{det_coeffs}) is the dispersion relation
\begin{equation}
\label{disp_rel_first}
e^{2\bar\varphi_+}(\bar{\cal A}^{\mu\nu}_+ k_\mu k_\nu)^2 \bar{\cal A}^{\sigma\omega}_- k_\sigma k_\omega - \epsilon^2\bigg[\frac{k\cdot k}{48\pi}(a^2\bar\alpha + b^2\bar\varepsilon) - \frac{ab}{12\pi} \bar\beta k\,\bar\zeta k\bigg] = {\cal O}(\epsilon^6)
\end{equation}
where $\bar {\cal A}^{\mu\nu}_+ k_\mu k_\nu = s\sqrt{\bar\alpha\bar\varepsilon}\,k\cdot k + 2s\bar\beta k\,\bar\zeta k$ and $\bar {\cal A}^{\mu\nu}_- k_\mu k_\nu = s\sqrt{\bar\alpha\bar\varepsilon}\,k\cdot k - 2s\bar\beta k\,\bar\zeta k$ have been used. Introducing the substitution $2\bar\beta k\,\bar\zeta k =  s\bar {\cal A}^{\mu\nu}_+ k_\mu k_\nu - \sqrt{\bar\alpha\bar\varepsilon}\, k\cdot k$ in the final term of (\ref{disp_rel_first}), and introducing $\bar {\cal A}^{\mu\nu}_- k_\mu k_\nu = 2s\sqrt{\bar\alpha\bar\varepsilon}\,k\cdot k -  \bar {\cal A}^{\mu\nu}_+ k_\mu k_\nu$ in the first term, yields
\begin{equation}
\label{disp_rel_+_sign}
e^{2\bar\varphi_+}(\bar {\cal A}^{\mu\nu}_+ k_\mu k_\nu)^2 + s\epsilon^2 \frac{1}{96\pi}\bigg[a \bigg(\frac{\bar\alpha}{\bar\varepsilon}\bigg)^{\frac{1}{4}} - b \bigg(\frac{\bar\varepsilon}{\bar\alpha}\bigg)^{\frac{1}{4}}\bigg]^2={\cal O}(\epsilon^3)
\end{equation}
since $\bar{\alpha}, \bar{\varepsilon} < 0$ and $\bar {\cal A}^{\mu\nu}_+ k_\mu k_\nu = {\cal O}(\epsilon)$. The latter follows from $\bar g^{\mu\nu}_+ k_\mu k_\nu = {\cal O}(\epsilon)$.

Inspection of (\ref{disp_rel_+_sign}) reveals the importance of the sign of $s$. Recall that the special case (\ref{quadratic_action_two_comp_special}) considered in Section~\ref{sec:quantum_theory} led to $s=-1$; hence, inspection of (\ref{disp_rel_+_sign}) shows that the covector $k_\mu$ is real. A complex $k_\mu$ would indicate losses whose physical origin would be unclear, so the fact that $k_\mu$ is real is reassuring. The quantum fluctuations of a uniform laser-driven plasma are not expected to produce real excitations in a perturbative context. Setting $s=-1$ and substituting $a$, $b$ using (\ref{ab_def}) leads to
\begin{equation}
\label{full_disp_rel_+}
\big|\sqrt{\bar\alpha \bar\varepsilon}\,k\cdot k + 2\bar\beta k\,\bar\zeta k\big| = \epsilon \frac{3}{4}\frac{1}{\sqrt{6\pi}}\frac{|k\cdot k|}{\big[(\sqrt{\bar\alpha \bar\varepsilon}+\bar\beta \cdot \bar\zeta)^2 - \bar\beta\cdot \bar\beta\,\bar\zeta\cdot\bar\zeta\big]^{1/4}}\bigg|\bigg(\frac{\bar\alpha}{\bar\varepsilon}\bigg)^{\frac{1}{4}} \bar\zeta k - \bigg(\frac{\bar\varepsilon}{\bar\alpha}\bigg)^{\frac{1}{4}} \bar\beta k\bigg| + {\cal O}(\epsilon^2)
\end{equation}
where $e^{-4\bar\varphi_+} = (\sqrt{\bar\alpha \bar\varepsilon}+\bar\beta \cdot \bar\zeta)^2 - \bar\beta\cdot \bar\beta\,\bar\zeta\cdot\bar\zeta$ has been used to eliminate $\bar\varphi_+$, which follows because $e^{-4\bar\varphi_+}$ is equal to the determinant of the tensor $\bar{A}_+{ }^\nu{ }_\mu = \eta_{\mu\sigma}\bar{\cal A}_+^{\sigma\nu}$.

The equivalent calculation using (\ref{Phi_lin_eq_dots_minus}), (\ref{Psi_lin_eq_dots_minus}) instead of (\ref{Phi_lin_eq_dots_plus}), (\ref{Psi_lin_eq_dots_plus}) yields
\begin{equation}
\label{full_disp_rel_-}
\big|\sqrt{\bar\alpha \bar\varepsilon}\,k\cdot k - 2\bar\beta k\,\bar\zeta k\big| = \epsilon \frac{3}{4}\frac{1}{\sqrt{6\pi}}\frac{|k\cdot k|}{\big[(\sqrt{\bar\alpha \bar\varepsilon}-\bar\beta \cdot \bar\zeta)^2 - \bar\beta\cdot \bar\beta\,\bar\zeta\cdot\bar\zeta\big]^{1/4}}\bigg|\bigg(\frac{\bar\alpha}{\bar\varepsilon}\bigg)^{\frac{1}{4}} \bar\zeta k + \bigg(\frac{\bar\varepsilon}{\bar\alpha}\bigg)^{\frac{1}{4}} \bar\beta k\bigg| + {\cal O}(\epsilon^2)
\end{equation}
since $\bar{\cal A}^{\mu\nu}_- k_\mu k_\mu = {\cal O}(\epsilon)$, and $e^{-4\bar\varphi_-} = (\sqrt{\bar\alpha \bar\varepsilon}-\bar\beta \cdot \bar\zeta)^2 - \bar\beta\cdot \bar\beta\,\bar\zeta\cdot\bar\zeta$ has been used to eliminate $\bar\varphi_-$. The latter is obtained because  $e^{-4\bar\varphi_-}$ is equal to the determinant of the tensor $\bar{A}_-{ }^\nu{ }_\mu = \eta_{\mu\sigma}\bar{\cal A}_-^{\sigma\nu}$. Finally, (\ref{Lorentzian_condition}) yields the condition
\begin{equation}
\label{Lorentzian_condition_bar}
-\bar\beta\cdot\bar\zeta > \sqrt{\bar\alpha\bar\varepsilon} + \sqrt{\bar\beta\cdot\bar\beta\,\bar\zeta\cdot\bar\zeta}
\end{equation}
on the unperturbed fields. In  (\ref{full_disp_rel_+}), (\ref{full_disp_rel_-}), (\ref{Lorentzian_condition_bar}) and throughout the following, a dot denotes the scalar product of a pair of vectors (or a pair of co-vectors) given by the spacetime metric $\eta_{\mu\nu}$ (or its inverse $\eta^{\mu\nu}$). 

The laser frequency is the highest of the frequencies contained in the unperturbed configuration, so it is natural to investigate (\ref{full_disp_rel_+}), (\ref{full_disp_rel_-}) in the context of an ultrarelativistic approximation for $\bar\beta^\mu$. We introduce the decomposition
\begin{equation}
\label{ultrarel_beta}
\bar\beta^\mu = \frac{1}{\check\epsilon}\bar\beta^\mu_{[-1]} + \check\epsilon\bar\beta^\mu_{[1]}
\end{equation}
of the timelike vector $\bar\beta^\mu$, where $\bar\beta^\mu_{[-1]}$, $\bar\beta^\mu_{[1]}$ are null vectors and $\check\epsilon$ is a perturbation parameter. The subscript enclosed by square parentheses in each coefficient in the decomposition (\ref{ultrarel_beta}) denotes the $\check\epsilon$-order of the term containing the coefficient, and we will use this convention throughout the following. Like $\epsilon$, the positive parameter $\check\epsilon$ has no intrinsic physical meaning; it is introduced solely to facilitate a perturbative expansion when $\check\epsilon\ll 1$, and it can be set to unity at the end of the calculation. Note $\bar\alpha=\bar\beta\cdot\bar\beta = 2\bar\beta_{[-1]}\cdot\bar\beta_{[1]}$ is non-zero and independent of $\check\epsilon$, and the inequality (\ref{Lorentzian_condition_bar}) is automatically satisfied to lowest order in $\check\epsilon$ because $\bar\beta_{[-1]}\cdot\bar\zeta < 0$, $\bar\varepsilon=\bar\zeta\cdot\bar\zeta+1$ and $\bar\zeta$ is independent of $\check\epsilon$.

To progress the analysis, it is fruitful to correlate $\check\epsilon$ with $\epsilon$ and express (\ref{full_disp_rel_+}), (\ref{full_disp_rel_-}) in terms of a single continuous parameter. Inspection of the result of inserting (\ref{ultrarel_beta}) into (\ref{full_disp_rel_+}), (\ref{full_disp_rel_-}) suggests the substitution $\epsilon=\sqrt{\check\epsilon^{2p-1}}$ where $p$ is a positive integer. Note $p\gg 1$ because the quantum corrections should be much smaller than the deviation of $\bar\beta^\mu$ from a null vector. Equations (\ref{full_disp_rel_+}), (\ref{full_disp_rel_-}) yield
\begin{equation}
\label{simp_disp_rel_+}
\big|\check\epsilon \sqrt{\bar\alpha\bar\varepsilon}\,k\cdot k + 2(\bar\beta_{[-1]}k + \check\epsilon^2\bar\beta_{[1]}k)\bar\zeta k\big| = \check\epsilon^p \frac{3}{4} \frac{1}{\sqrt{6\pi}}\frac{|k\cdot k|}{\sqrt{|\bar\beta_{[-1]}\cdot\bar\zeta|}}\bigg(\frac{\bar\varepsilon}{\bar\alpha}\bigg)^{\frac{1}{4}} \big|\bar\beta_{[-1]}k\big| + {\cal O}(\check\epsilon^{p+1})
\end{equation}
and
\begin{equation}
\label{simp_disp_rel_-}
\big|\check\epsilon \sqrt{\bar\alpha\bar\varepsilon}\,k\cdot k - 2(\bar\beta_{[-1]}k + \check\epsilon^2\bar\beta_{[1]}k)\bar\zeta k\big| = \check\epsilon^p \frac{3}{4} \frac{1}{\sqrt{6\pi}}\frac{|k\cdot k|}{\sqrt{|\bar\beta_{[-1]}\cdot\bar\zeta|}}\bigg(\frac{\bar\varepsilon}{\bar\alpha}\bigg)^{\frac{1}{4}} \big|\bar\beta_{[-1]}k\big| + {\cal O}(\check\epsilon^{p+1})
\end{equation}
respectively, both of which give
\begin{equation}
\label{zero_order_disp}
\bar\beta_{[-1]}k_{[0]}\,\bar\zeta k_{[0]} = 0
\end{equation}
where $k_\mu = k_{[0]\mu} + \check\epsilon k_{[1]\mu} + {\cal O}(\check\epsilon^2)$ has been introduced. Equation (\ref{zero_order_disp}) can be solved by $\bar\beta_{[-1]}k_{[0]} = 0$ or $\bar\zeta k_{[0]} = 0$. If $\bar\beta_{[-1]}k_{[0]} = 0$ then, since $\bar\beta^\mu_{[-1]}$ is null, it follows $k^\mu_{[0]}$ is proportional to $\bar\beta^\mu_{[-1]}$, where $k^\mu_{[0]} = \eta^{\mu\nu}k_{[0]\nu}$. Alternatively, if $\bar\zeta k_{[0]} = 0$ then $k^\mu_{[0]}$ must be spacelike because $\bar\zeta^\mu$ is timelike. The quantity $\bar\zeta k$ is proportional to the frequency of the perturbations $\Phi_{(1)}$, $\Psi_{(1)}$ in the rest frame of the plasma electrons, and if $\bar\zeta k_{[0]} = 0$ then this frequency vanishes to lowest order in $\check\epsilon$.

Since $p>1$ the result $\bar\beta_{[-1]}k_{[1]}\,\bar\zeta k_{[0]} = 0$ immediately emerges from both (\ref{simp_disp_rel_+}), (\ref{simp_disp_rel_-}) when $\bar\beta_{[-1]}k_{[0]}=0$; thus, $k_{[1]\mu}$ is proportional to $k_{[0]\mu}$ in this case. Up to first order in $\check\epsilon$, the phase speed of the perturbations $\Phi_{(1)}$, $\Psi_{(1)}$ is the speed of light in vacuo. However, the behaviour of the first order term when $\bar\zeta k_{[0]}=0$ is quite different. In this case, it is useful to decompose the contractions within (\ref{simp_disp_rel_+}), (\ref{simp_disp_rel_-}) with respect to the timelike unit vector $n^\mu = \bar\zeta^\mu/\sqrt{-\bar\zeta\cdot\bar\zeta}$ and a spacelike unit vector $n_\perp^\mu$ orthogonal to $n^\mu$. We find
\begin{equation}
\label{nk_first_order}
n k_{[1]} = - \sqrt{\frac{\bar\alpha\bar\varepsilon}{-\bar\zeta\cdot\bar\zeta}}\,\frac{n_\perp k_{[0]}}{2 n_\perp\cdot\bar\beta_{[-1]}},\qquad 
n k_{[1]} = \sqrt{\frac{\bar\alpha\bar\varepsilon}{-\bar\zeta\cdot\bar\zeta}}\,\frac{n_\perp k_{[0]}}{2 n_\perp\cdot\bar\beta_{[-1]}}
\end{equation}
emerges from (\ref{simp_disp_rel_+}), (\ref{simp_disp_rel_-}) respectively, where $\eta^{\mu\nu} = - n^\mu n^\nu + n_\perp^\mu n_\perp^\nu$ has been used. Note that $n_\perp k_{[0]}$ is proportional to the wavenumber of the perturbations $\Phi_{(1)}$, $\Psi_{(1)}$ in the rest frame of the plasma electrons and so, up to first order in $\check\epsilon$, a wave packet formed from those perturbations will propagate without dispersing. In fact, this statement holds up to $(p-1)$th order in $\check\epsilon$ because, up to that order, the contents of the modulus brackets on the left-hand sides of (\ref{simp_disp_rel_+}), (\ref{simp_disp_rel_-}) are second-order homogenous polynomials in $k_\mu$. The wave packet will not disperse without the contribution of the right-hand sides of (\ref{simp_disp_rel_+}), (\ref{simp_disp_rel_-}).

Corrections due to quantum fluctuations only contribute at $p$th order and above in $\check\epsilon$. However, inspection of (\ref{simp_disp_rel_+}), (\ref{simp_disp_rel_-}) shows that their right-hand sides  are ${\cal O}(\check\epsilon^{p+1})$ when $k^\mu_{[0]}$ is proportional to $\bar\beta^\mu_{[-1]}$, so the quantum corrections are insignificant in this case. However, analysis of the case where $n k_{[0]} = 0$ reveals
\begin{equation}
\label{ultrarel_disp_simp}
n k = \pm \nu\,n_\perp k - \check\epsilon^p \frac{3}{8} \frac{1}{\sqrt{6\pi}} \frac{1}{\sqrt{|n\cdot\bar\beta_{[-1]}|}} \bigg(\frac{\bar\varepsilon}{|\bar\zeta\cdot\bar\zeta|^3\bar\alpha}\bigg)^{\frac{1}{4}} (n_\perp k)^2 + {\cal O}(\check\epsilon^{p+1})
\end{equation}
where the negative sign corresponds to (\ref{simp_disp_rel_+}), the positive sign corresponds to (\ref{simp_disp_rel_-}), and the sign of the quantum correction has been chosen so that its contribution to the frequency of $\Phi_{(1)}$, $\Psi_{(1)}$ is positive for all $n_\perp k$. Likewise, the sign of the first term in (\ref{ultrarel_disp_simp}) is fixed by requiring that it makes a positive contribution to the frequency of $\Phi_{(1)}$, $\Psi_{(1)}$ for each $n_\perp k$. The constant $\nu$ is
\begin{equation}
\label{phase_vel_disp}
\nu = \sqrt{\frac{\bar\alpha\bar\varepsilon}{-\bar\zeta\cdot\bar\zeta}}\,\frac{\check\epsilon}{2 n_\perp\cdot\bar\beta_{[-1]}} + {\cal O}(\check\epsilon^2).
\end{equation}

Finally, we will now express (\ref{ultrarel_disp_simp}), (\ref{phase_vel_disp}) in terms of the dimensionful variables that were introduced in the context of the underlying $3$-dimensional classical theory in Section~\ref{sec:classical_theory}. Dropping ${\cal O}(\check\epsilon^{p+1})$ from (\ref{ultrarel_disp_simp}), and setting $\check\epsilon$ to unity, yields the dispersion relation
\begin{equation}
\label{ultrarel_disp_final}
\omega = v\kappa + \frac{3}{8}\sqrt{\frac{\hbar e^2}{6\pi\varepsilon_0 m_e^2 c^3 L_*^2}}\,\bigg[\frac{a_0^2}{(a_0^2+1)^3}\bigg]^{\frac{1}{4}}\frac{c^2 \kappa^2}{\sqrt{\omega_0 \omega_p}}\bigg|_{x=y=0}
\end{equation}
where the speed $v$ is
\begin{equation}
\label{ultrarel_phase_speed}
v = \frac{\omega_p}{\omega_0} \frac{a_0 c}{2\sqrt{a_0^2+1}}\bigg|_{x=y=0} + {\cal O}(\omega_0^{-2}).
\end{equation}
The angular frequency $\omega$ and wavenumber $\kappa$ of the perturbation in the rest frame of the plasma electrons are given by $\omega = c|n k|/l_*$ and $\kappa = |n_\perp k|/l_*$, respectively, where $l_*$ is the inert length scale used in the construction of (\ref{1+1_action}). The laser frequency $\omega_0$, laser wavenumber $k_0$, plasma frequency $\omega_p$, and dimensionless laser amplitude
\begin{equation}
a_0 = \frac{e\sqrt{\langle{\bf A}_0^2\rangle}}{m_e c}
\end{equation}
emerge using the substitutions
\begin{align}
\label{final_substitutions_beta}
&-n\cdot\bar\beta_{[-1]} = |n_\perp\cdot\bar\beta_{[-1]}| = \sqrt{\frac{\varepsilon_0 m_e^2 c^3 L_*^2}{\hbar e^2}}\frac{l_*\,\omega_0}{c}\bigg|_{x=y=0},\\
\label{final_substitutions_alpha_zeta}
&\bar\zeta\cdot\bar\zeta = \bar\varepsilon - 1 = -(a_0^2 + 1)\big|_{x=y=0},\quad \bar\alpha = -\frac{\varepsilon_0 m_e^2 c^3 L_*^2}{\hbar e^2}\frac{l_*^2\,\omega_p^2}{c^2}\bigg|_{x=y=0}.
\end{align}
The details of (\ref{final_substitutions_beta}), (\ref{final_substitutions_alpha_zeta}) follow from (\ref{dimensionless_variables}), $\beta_\mu = \partial_\mu \Phi$, $\zeta_\mu = \partial_\mu \Psi$,
\begin{equation}
(\partial_t\widetilde\Phi)^2 - c^2(\bm{\nabla}\widetilde\Phi)^2 = \omega_p^2,\qquad (\partial_t\widetilde\Psi)^2 - c^2(\bm{\nabla}\widetilde\Psi)^2 - m_e^2 c^4 = e^2 c^2 \langle {\bf A}_0^2\rangle,
\end{equation}
and $\partial_x \widetilde{\Phi}|_{x=y=0} = \partial_y \widetilde{\Phi}|_{x=y=0} = 0$, $\partial_x \widetilde{\Psi}|_{x=y=0} = \partial_y \widetilde{\Psi}|_{x=y=0} = 0$. Equation (\ref{ultrarel_disp_final_intro}) follows immediately from (\ref{ultrarel_disp_final}) upon replacing $L_*$ with the width of the laser beam.

In summary, we have identified two distinct dispersion relations, in the regime $\omega_0 \gg \omega_p$, describing dynamical perturbations of a uniform underdense laser-driven plasma. One of the dispersion relations describes propagation in the same direction, at essentially the same phase speed, as the laser beam. The remaining dispersion relation is associated with perturbations that co-propagate and counter-propagate with the laser beam, but at a much slower speed than the laser beam. None of the modes are dispersive without quantum corrections, and the modes that propagate at essentially the same speed as the laser beam are non-dispersive even when quantum effects are included. The behaviour of the slow modes is given by (\ref{ultrarel_disp_final}), where $\kappa$ is the magnitude of the wave vector.
\appendix
\section{Relationship between the bi-scalar field theory and scalar QED}
\label{appendix:QED}
For notational simplicity, we will use natural units throughout the following. Consider the Lagrangian
\begin{equation}
\label{lagrangian_QED_4d}
{\cal L}_{(4)} = -\frac{1}{2}D^a\Xi^*\, D_a \Xi - \frac{1}{2}m^2 |\Xi|^2 - \frac{1}{4} F^{ab}F_{ab} - A^a J_a
\end{equation}
for an electromagnetic $4$-potential $A_a$ and a complex scalar field $\Xi$ with mass $m=m_e$ and electromagnetic charge $q=-e$. The vector field $J^a$ is the electric $4$-current of the ion background. The $U(1)$-covariant derivative is $D_a = \partial_a - iq A_a$, and $F_{ab} = \partial_a A_b - \partial_b A_a$. Lowercase Latin indices $a,b$ range over $0, 1, 2, 3$, and indices are lowered and raised using the $4$-dimensional Minkowski spacetime metric $\eta^{(4)}_{ab}$ and its inverse $\eta_{(4)}^{ab}$, respectively. The metric $\eta^{(4)}_{ab}$ has signature $(-,+,+,+)$.

We will now argue that (\ref{1+1_action}) emerges from a dimensionally-reduced theory induced from (\ref{lagrangian_QED_4d}). The flow of energy and momentum is predominantly along the $x^1$-axis of the Minkowski coordinate system $x^0, x^1, x^2, x^3$, so it is natural to expand the contractions in (\ref{lagrangian_QED_4d}) and neglect the dependence of the fields on the coordinates $x^2, x^3$. This procedure yields
\begin{equation}
\label{lagrangian_QED_2d}
{\cal L} = -\frac{1}{2}D^\mu\Xi^*\, D_\mu \Xi - \frac{1}{2}q^2|\mathbb{A}|^2|\Xi|^2 - \frac{1}{2}m^2 |\Xi|^2 - \frac{1}{4} F^{\mu\nu}F_{\mu\nu} - \frac{1}{2}\partial^\mu\mathbb{A}^*\partial_\mu\mathbb{A} - A^\mu J_\mu
\end{equation}
where the fields $\Xi, A_\mu$, and the complex field $\mathbb{A}=A_2+iA_3$, only depend on $x^0, x^1$. Furthermore, $J^2=J^3=0$ has been assumed. Finally, the action $\int d^4 x \sqrt{-\eta^{(4)}}{\cal L}_{(4)}$ for the $4$-dimensional theory is replaced by $L_*^2 \int d^2 x \sqrt{-\eta}{\cal L}$, where the length $L_*$ characterises the size of the domain in the $x^2-x^3$ plane on which (\ref{lagrangian_QED_4d}) is non-zero. 

Let $|{\rm I}\rangle \equiv |A_{\mu{\rm I}},\mathbb{A}_{\rm I}, \Xi_{\rm I}; x^0_{\rm I}\rangle$, $|{\rm II}\rangle \equiv |A_{\mu{\rm II}},  \mathbb{A}_{\rm II}, \Xi_{\rm II}; x^0_{\rm II}\rangle$ be eigenstates of the field operators $\hat{A}_{\mu}, \hat{\mathbb{A}}, \hat{\Xi}$ at time $x^0_{\rm I}$, $x^0_{\rm II}$, respectively. The transition amplitude between the two sets of fields is
\begin{equation}
\label{transition}
\langle{\rm II}|{\rm I}\rangle = \int {\cal D}A_\mu\,{\cal D}\mathbb{A}^*\,{\cal D}\mathbb{A}\,{\cal D}\Xi^*{\cal D}\Xi\, \exp\bigg(iL_*^2\int d^2 x \sqrt{-\eta}{\cal L}\bigg)
\end{equation}
where it is understood that the lower and upper integration limits in the action integral in (\ref{transition}) are $x^0_{\rm I}$, $x^0_{\rm II}$, respectively. To proceed, it is fruitful to introduce the polar forms $\Xi = |\Xi|\exp(i\Psi)$, $\mathbb{A} = |\mathbb{A}|\exp(i\Phi)$; thus
\begin{align}
\label{Xi_polar}
D^\mu \Xi^* D_\mu \Xi &= (\partial^\mu\Psi - q A^\mu)(\partial_\mu\Psi - q A_\mu)\,|\Xi|^2 + \partial^\mu |\Xi|\,\partial_\mu |\Xi|,\\
\label{A_polar}
\partial^\mu\mathbb{A}^*\partial_\mu\mathbb{A} &= \partial^\mu \Phi \partial_\mu \Phi\,|\mathbb{A}|^2 + \partial^\mu|\mathbb{A}|\partial_\mu|\mathbb{A}|
\end{align}
and
\begin{equation}
\label{transition_polar}
\langle{\rm II}|{\rm I}\rangle = \int {\cal D}A_\mu{\cal D}\Phi {\cal D}\Psi\,{\cal D}(|\mathbb{A}|^2) {\cal D}(|\Xi|^2)  \exp\bigg(iL_*^2\int d^2 x \sqrt{-\eta}{\cal L}\bigg)
\end{equation}
since ${\cal D}\mathbb{A}^*\,{\cal D}\mathbb{A} = 2|\mathbb{A}|{\cal D}|\mathbb{A}|\,{\cal D}\Phi$ and ${\cal D}\Xi^*\,{\cal D}\Xi = 2|\Xi|{\cal D}|\Xi|\,{\cal D}\Psi$.

The states $|{\rm I}\rangle$, $|{\rm II}\rangle$ are based on two instantaneous classical configurations of a laser-driven plasma in which the derivatives of the phases $\Psi$, $\Phi$ dominate over the derivatives of $|\Xi|$, $|\mathbb{A}|$. Furthermore, the dominant contributions to (\ref{transition_polar}) will be from field trajectories that are close to the classical trajectory connecting $A_{\mu{\rm I}},\mathbb{A}_{\rm I}, \Xi_{\rm I}$ with $A_{\mu{\rm II}},\mathbb{A}_{\rm II}, \Xi_{\rm II}$. Thus, in calculating (\ref{transition_polar}), it is reasonable to only integrate over field trajectories for which the kinetic terms $\partial^\mu |\Xi|\,\partial_\mu |\Xi|$, $\partial^\mu|\mathbb{A}|\partial_\mu|\mathbb{A}|$ are small relative to the remaining terms in the Lagrangian (\ref{lagrangian_QED_2d}).

Inspection of the remaining terms in (\ref{lagrangian_QED_2d}) shows that the integrals over $|\mathbb{A}|^2$, $|\Xi|^2$ in (\ref{transition_polar}) are infinite-dimensional analogues of
\begin{equation}
\label{exact_int}
I = \int^\infty_0 dx \int^\infty_0 dy\,e^{i(ax+by+cxy)}
\end{equation}
where $a$, $b$ have small positive imaginary parts, and $c$ is a negative real number. The properties of the imaginary parts of $a$, $b$ agree with the Feynman prescription ($\text{mass}^2 \mapsto \text{mass}^2 - i\varepsilon$, $\varepsilon>0$) for $\Xi$, $A_a$ in (\ref{lagrangian_QED_4d}). By evaluating either integral, it is easy to show 
\begin{equation}
I = \frac{i}{c}e^{-iab/c} \underset{\varepsilon\rightarrow 0^+}\lim\int^\infty_{{\rm Re}(\frac{a}{c})} dz\,\frac{e^{ibz}}{z-i\varepsilon} = \frac{i}{c}e^{-iab/c} \underset{\varepsilon\rightarrow 0^+}\lim\int^\infty_{{\rm Re}(\frac{b}{c})} dz\,\frac{e^{iaz}}{z-i\varepsilon}
\end{equation}
and, with $\theta$ denoting the Heaviside function, it follows
\begin{equation}
\label{approx_int}
I \approx -\frac{2\pi}{c} \theta(a)\theta(b)e^{-iab/c}
\end{equation}
in the limit ${\rm Im}(a), {\rm Im}(b) \rightarrow 0^+$, when $|a/c|$ and $|b/c|$ are large.

Equation (\ref{approx_int}) suggests that the functional integrals over $|\mathbb{A}|^2$, $|\Xi|^2$ in (\ref{transition_polar}) give
\begin{equation}
\label{transition_polar_approx}
\langle{\rm II}|{\rm I}\rangle \approx \int {\cal D}A_\mu{\cal D}\Phi {\cal D}\Psi\,\theta[U_\Phi]\theta[U_\Psi] \exp\bigg(iL_*^2 \int d^2 x \sqrt{-\eta} {\cal L}^\prime\bigg)
\end{equation}
up to an overall multiplicative constant, where
\begin{equation}
\label{lagrangian_reduced_2d}
{\cal L}^\prime =  \frac{1}{2q^2} \{(\partial^\mu\Psi - q A^\mu)(\partial_\mu\Psi - q A_\mu) +m^2\}\partial^\nu \Phi \partial_\nu \Phi - \frac{1}{4} F^{\mu\nu}F_{\mu\nu} - A^\mu J_\mu
\end{equation}
and
\begin{align}
U_\Phi = - \partial^\mu\Phi \partial_\mu\Phi,\qquad U_\Psi = - (\partial^\mu\Psi - q A^\mu)(\partial_\mu\Psi - q A_\mu) - m^2
\end{align}
with $\theta[\cdot]$ denoting the functional Heaviside. The fields used to evaluate (\ref{transition_polar}) exclude cases where $U_\Phi$, $U_\Psi$ pass through zero. In particular, the conditions
\begin{equation}
\label{U_conditions}
q^2\bigg|\frac{\partial^\mu \sqrt{|U_\Phi|}\partial_\mu\sqrt{|U_\Phi|}}{U_\Phi U_\Psi}\bigg| \ll 1,\qquad q^2\bigg|\frac{\partial^\mu \sqrt{|U_\Psi|}\partial_\mu\sqrt{|U_\Psi|}}{U_\Phi U_\Psi}\bigg| \ll 1
\end{equation}
are necessary because $\partial^\mu |\mathbb{A}| \partial_\mu |\mathbb{A}|$, $\partial^\mu |\Xi| \partial_\mu |\Xi|$ are small, and the action in (\ref{transition_polar}) is stationary when $q^2 |\Xi|^2 \approx U_\Phi$ and $q^2 |\mathbb{A}|^2 \approx U_\Psi$. 

Up to rescalings of $\Phi$, $\Psi$, $x^\mu$, the Lagrangian (\ref{lagrangian_reduced_2d}) is simply that of (\ref{1+1_action}) when the effects of $A_\mu$ are negligible. However, there is an important difference: unlike their brethren in (\ref{1+1_action}), the functions $\Phi$, $\Psi$ in (\ref{lagrangian_reduced_2d}) have finite ranges because they are angles in polar decompositions of $\mathbb{A}$, $\Xi$. Nevertheless, the two theories are essentially equivalent when$\sqrt{|U_\Psi|}$, $\sqrt{|U_\Phi|}$ are large.

Comparison of the first term in (\ref{lagrangian_reduced_2d}) with the Lagrangian for a free non-relativistic particle confined to a $1$-dimensional ring suggests that $\sqrt{|U_\Psi|}$, $\sqrt{|U_\Phi|}$ are each analogous to the radius of the ring, with $\Phi$, $\Psi$ the corresponding angles. Small time-dependent perturbations to the angle of the particle around the ring will make substantial contributions to the Lagrangian of the particle when the radius of the ring is large. Likewise, the response of (\ref{lagrangian_reduced_2d}) to small changes in $\Phi$, $\Psi$ will be large when $\sqrt{|U_\Psi|}$, $\sqrt{|U_\Phi|}$ are large. The aggressiveness of the destructive interference of the non-classical contributions to the transition amplitude (\ref{transition_polar_approx}) increases with $\sqrt{|U_\Psi|}$, $\sqrt{|U_\Phi|}$. Hence, the fact that $\Phi$, $\Psi$ have finite ranges lessens in physical significance with increasing $\sqrt{|U_\Psi|}$, $\sqrt{|U_\Phi|}$. In this case, there are no significant physical consequences of extending the ranges of $\Phi$, $\Psi$ in (\ref{transition_polar_approx}) to the entire real line. Thus, it is reasonable to treat $\Phi$, $\Psi$ as bona fide scalar fields when $\sqrt{|U_\Psi|}$, $\sqrt{|U_\Phi|}$ are large, and it is natural to use the perturbative approach (\ref{1-loop_action}) to calculate the quantum corrections to the classical field equations of (\ref{lagrangian_reduced_2d}). To keep the calculation tractable, the quantum corrections in (\ref{1-loop_action}) are evaluated in the limit $x^0_{\rm I} \rightarrow -\infty$, $x^0_{\rm II} \rightarrow \infty$, although the domain of integration in the classical action in (\ref{1-loop_action}) is finite in time.

Finally, since the kinetic terms of $|\Xi|$, $|\mathbb{A}|$ are negligible throughout the above procedure, our approach does not capture the effects of quantum (in particular, zero-point) fluctuations of $|\Xi|$, $|\mathbb{A}|$ about their classical values. The effective action calculated here includes the quantum fluctuations of $\Phi$, $\Psi$ only.
\section*{Acknowledgements}
We thank Robin W Tucker, Hartmut Ruhl, and Daniel Seipt for useful discussions, and we thank the anonymous referees for their useful comments. This work was supported by the UK Engineering and Physical Sciences Research Council grant EP/N028694/1 (D.A.B., A.C., A.N.), the Cockcroft Institute (A.C.), and the Lancaster University Faculty of Science and Technology (C.F.). A.C. is supported by the Irish Research Council Postdoctoral Fellowship GOIPD/2019/536. All of the results can be fully reproduced using the methods described in the paper.

\end{document}